\documentclass[useAMS,usenatbib]{mn2e}

\usepackage{amsmath}
\usepackage{graphics}
\usepackage{graphicx}

\def\lcdm{$\Lambda$CDM }

\newcommand{\hMpc}{\mbox{$h^{-1}$ Mpc} }

\newcommand{\hMpcKet}{\mbox{$h^{-1}$ Mpc)} }

\newcommand{\hMpcCom}{\mbox{$h^{-1}$ Mpc,} }
\newcommand{\hMpcDot}{\mbox{$h^{-1}$ Mpc.} }

\newcommand{\hmsun}{\mbox{$h^{-1}$ $M_{\odot}$} }

\def\la{\mathrel{\hbox{\rlap{\hbox{\lower4pt\hbox{$\sim$}}}\hbox{$<$}}}}
\def\ga{\mathrel{\hbox{\rlap{\hbox{\lower4pt\hbox{$\sim$}}}\hbox{$>$}}}}

\def\gsim{\ga}

\newcommand{\bc}{\begin{center}}
\newcommand{\ec}{\end{center}}

\newcommand{\be}{\begin{equation}}
\newcommand{\ee}{\end{equation}}

\newcommand{\tx}[1] {\rmn{#1}}

\title[Reconstruction of primordial density fields]
{Reconstruction of primordial density fields}
\author[Roya Mohayaee, Hugues Mathis, St\'ephane Colombi and Joseph Silk]
{Roya Mohayaee$^*$,  Hugues Mathis$^\dagger$, 
St\'ephane Colombi$^*$ and Joseph Silk$^\dagger$ \\
$^*$Institut d'Astrophysique de Paris, 98bis boulevard Arago, 75014 Paris, France\\ 
$^\dagger$University of Oxford, Astrophysics, Denys
Wilkinson Building, Keble Road, Oxford OX1 3RH, UK}

\begin{document}
\maketitle
\begin{abstract}
The Monge-Amp\`ere-Kantorovich (MAK) reconstruction is tested
against cosmological $N$-body simulations. Using only the present 
mass distribution sampled with particles, 
and the assumption of homogeneity of the primordial 
distribution, MAK recovers for each particle 
the non-linear displacement field between its present position 
and its Lagrangian position on a primordial uniform grid. 

To test the method, 
we examine a standard $\Lambda$CDM $N$-body simulation 
with Gaussian initial conditions and 6 models 
with non-Gaussian initial conditions: a $\chi^2$ model, a model with
primordial voids and four weakly non-Gaussian models.

Our extensive analyses of the Gaussian simulation
show that the level of accuracy of the reconstruction of the nonlinear
displacement field achieved by MAK is unprecedented, at scales
as small as $\sim 3\hMpcDot$ In particular, it captures in a
nontrivial way the nonlinear contribution from gravitational instability, well
beyond the Zel'dovich approximation. This is also confirmed by our
analyses of the non-Gaussian samples. 

Applying the spherical collapse model to the probability distribution function
of the divergence of the displacement field, we also show that 
from a well-reconstructed displacement field, such as that given
by MAK, it is possible to accurately disentangle
dynamical contributions induced by gravitational clustering 
from possible initial non-Gaussianities, allowing one to
efficiently test the non-Gaussian nature of the primordial
fluctuations.

In addition, we test successfully
a simple application of MAK using the Zel'dovich approximation
to recover in real space the present-day peculiar velocity field on scales of
$8\ h^{-1}$ Mpc.

Although non trivial observational issues yet remain to be addressed,
our numerical investigations suggest that MAK reconstruction represents a 
very promising tool to be applied to three dimensional galaxy catalogs.
\end{abstract}

\begin{keywords}  
cosmology: dark matter -- large-scale structure of the Universe
methods: $N$-body simulations -- numerical 
        \end{keywords} 



\section[]{Introduction}
\label{sec:Intro}

Recent data from the Cosmic Microwave 
Background (CMB), from the redshift surveys of the
distribution of galaxies on large scales, and from weak 
lensing surveys on the projected cosmic
density field have allowed not only a precise determination of the main
cosmological parameters, but they have also confirmed 
the consistency and validity of
the cold dark matter (CDM) paradigm for structure formation. The
values of cosmological parameters have been pinpointed
by the \emph{WMAP} data \citep{Spe03} from the power spectrum of 
the CMB temperature fluctuations, by
a variety of results from the 2dFGRS and the SDSS such as the degree of anisotropy of the
redshift space correlation of bright galaxies or their degree 
of clustering (respectively, \citealt{Peac01,Teg03a}),
or by lensing surveys (for example, \citealt{Wae01,Pen03}).

Some of these results depend on priors assumed for the primordial matter 
power spectrum (for example a ``tilt'' or a running index),
and also on the nature of the dark matter, which affects the transfer 
function. However, recent measurements of both the
present-day density correlations and the CMB fluctuations are consistent
 with the gravitational instability theory. The theory asserts that the structures we observe
 today have evolved from infinitesimal, adiabatic, Gaussian, scale-invariant primordial
density fluctuations generated  in minimal models of inflation \citep{Pei03,Kom03,Teg03b},
 and indicate that dark matter is ``cold'' at least at scales greater than about 100 kpc.

Various details of the paradigm still need to be 
constrained. An important aspect is the shape of 
the probability distribution function (PDF) of the primordial
matter density field and in particular its possible deviations from Gaussianity on scales
relevant to CMB and/or to large-scale structure (hereafter LSS) of the galaxy distribution
(see \citealt{Pee83} for early motivations). In fact, the simplest 
single-field inflationary models
have been shown to be able to produce only a very small amount of non-Gaussianity below
that detectable by current experiments (for example, \citealt{Malda03,Acq03}).
However, a variety of physically plausible models, including those where the
seeds for structure formation, at least partly, takes the form of topological
defects (\citealt{Ave98}  and references therein) or those associated
to multifield inflation (e.g. \citealt{Lin97,Pee99a, Bern02}) can 
produce significant relic non-Gaussianity (see also \citealt{Bar04}).

Recently, constraints on primordial CMB-scale non-Gaussianity have been 
obtained from higher-order statistics of
temperature CMB maps (for example \citealt{Kom03}) assuming a 
parametrisation which bears the form of a
quadratic nonlinearity in the primordial gravitational potential. Additional 
limits are provided by the abundance
of massive clusters both today and at moderately high redshifts \citep{Chi98,Koy99,RGS00,Mat00}.

Although current CMB maps and cluster abundances are consistent with 
primordial fluctuations being Gaussian,
certain subtleties should be kept in mind. First, the sensitivity of a given 
test in detecting primordial non-Gaussianity
depends on the parametrisation chosen for the deviations from 
Gaussianity (e.g. whether one describes
deviations in the gravitational potential or directly in the density:
\citealt{Ver01}). Secondly, at a fixed comoving scale,  Gaussianizing 
or non-Gaussianizing biases that can be induced by data
reduction and also by gravitational clustering, if the scale is 
small, need to be carefully assessed and controlled. Finally,
non-Gaussianity might be strongly scale-dependent as for models
including topological defects or other kinds of phase transitions in the early
Universe.  Therefore, as is the case for the determination 
of cosmological parameters, the measurement of primordial
non-Gaussianity will also benefit from combining different methods which 
rely on different assumptions, are sensitive
to different shapes of non-Gaussianity and probe different scales. Assessing 
the possibilities of an original
method in obtaining the statistics of the primordial density field especially 
on comoving scales of $\sim 3$ to $\gsim 10$ \hMpc
is the subject of this paper.

A variety of methods which use LSS rather than CMB for 
the recovery of the statistics
of the primordial density field have been developed recently, some of which
rely on measurement of  higher-order correlations
of the galaxy distribution \citep{Ver01,Fel01,Sco04} and some others on the statistics 
of the cosmic shear alone \citep{Taka04} or in combination with the bright X-ray cluster 
abundances \citep{Am03}. The main difficulty with
extracting primordial non-Gaussianity out of the statistics of the present-day
LSS resides in the contribution to non-Gaussianity at lower redshifts
by the nonlinear gravitational evolution of the density field \citep{Juszk93}:
gravity can generate skewness and higher-order moments 
in a random, initially Gaussian density
field \citep{Pee80,Bouch92}.

An alternative approach for testing non-Gaussianity, consists of 
evolving the present time density field
inferred from the galaxy catalogues back in time to higher redshifts, using
reconstruction methods.
Attempts to detect primordial non-Gaussianity applying
reconstruction methods to galaxy surveys have been made, for example using the 1.2 Jy \emph{IRAS}
survey (\citealt{Nus95}) or \emph{IRAS} PSC redshift catalogue \citep{Mon00}. 
Under the assumption of a linear bias the statistics of the primordial 
density field were found to be consistent with
Gaussianity \citep{Mon00}. However, most of these reconstruction 
methods require a significant smoothing
of the evolved density field to make it linear.
This restriction calls for a Lagrangian based
reconstruction algorithm which could be applicable at smaller scales.

Variational methods have already been extensively used for the reconstruction of the primordial
density and the present peculiar velocity fields, the determination of masses of galaxies and
clusters and of the cosmological parameters. In these methods 
one starts from the present positions
of the galaxies taken as mass tracers or of the dark matter particles (in the 
case of simulations) and finds their peculiar velocities
and subsequently their full orbits.  These peculiar velocities 
(or orbits) can be obtained by minimising the Euler-Lagrange
action \citep{Pee89a,Shay95,Nus00}, by using the perturbative Eulerian or Lagrangian theories
\citep{Gram93,Nus95,Mon00} or by solving an optimal assignment problem
with stochastic algorithms (\citealt{Cro97}) or deterministic algorithms, such
as the Monge-Amp\`ere-Kantorovitch
(MAK) assignment method (\citealt{Frisch02,Moh03,Bre03}, hereafter FMB) which is used
in this work. 

A difficulty with the reconstructions based on Peebles' action 
modeling is their lack of uniqueness:  given
a present distribution of mass, many different primordial
 density fields and past histories, all physically plausible, can be
 reconstructed. All of these solutions correspond to stationary points
 of the action. This can be a serious drawback when the primordial positions for several
million particles have to be reconstructed, because
 the number of solutions can increase accordingly.
An additional difficulty related to the lack of uniqueness of the solution
is that variational-based methods cannot guarantee that the full possible solution space has
been explored. In addition, fast least-action-based algorithms which could be
realistically applied to large datasets do not exist whereas
fast MAK algorithms have been developed and are applied for the first
time here to datasets of the order of $2$ million particles.
Furthermore, it has been shown that MAK reconstruction 
is a well-defined problem at large comoving scales
($\sim$ a few \hMpcKet where multi-steaming 
can be neglected and has a unique solution (see FMB and references
therein).

In this work, we test the MAK reconstruction method against $N$-body simulations.
We first focus on a simulation with Gaussian initial condition, namely
standard $\Lambda$CDM model. We check the ability of MAK as   
a reconstructor of the non-linear displacement field, which is
the difference between the present position of a particle in the simulation 
and its initial position on a uniform grid.
We extend the analyses to simulations with non-Gaussian initial conditions,
and check whether MAK can detect the primordial non-Gaussianities.
The non-Gaussian models examined in this work are
a $\chi^2$ model where
the density field is the square of a Gaussian as a proxy 
for a strong non-Gaussianity which is scale-independent in our
simulated volume, a primordial voids model ({\it PVM}) to depict a scale-dependent non-Gaussianity
which is significant at small scales but negligible on large scales, and 4
models with quadratic corrections to the primordial 
gravitational potential ($Q$ models) to represent a minute scale-independent non-Gaussianity 
consistent with the CMB (see for example
\citealt{Kom03}). We finally interpret some of the results of the analyses
using the spherical collapse model.

The paper is organised as follows. Section~\ref{sec:Models} summarises the features of
the Gaussian and non-Gaussian models studied here and describes the setup of 
the corresponding $N$-body simulations.
An outline of our reconstruction method is provided in 
Section~\ref{sec:Method}. Section~\ref{sec:ResultsGauss}
deals with the reconstruction of the Gaussian model. We demonstrate 
that MAK is an extremely good
reconstructor of the non-linear displacement field measured in the 
simulation, despite
possible limitations due to shell crossing. 
Section~\ref{sec:ResultsNonGauss} is devoted to the application of MAK to our series 
of plausible non-Gaussian models.
Since MAK is a nonlinear reconstructor, we find that the primordial
non-Gaussianities are mixed up with non-Gaussianities 
developed by gravitational instability. A simple but efficient 
way of disentangling these
two non-Gaussianities is illustrated in
Section~\ref{sec:Prospects} in terms of the spherical collapse model.
In Section~\ref{sec:CCL}, we summarise the main results
of this paper and discuss the potential extra complications 
that could arise in the application of MAK to real galaxy catalogues.


\section{Simulating models of the primordial density field}
\label{sec:Models}

We first describe the general features common 
to our 7 simulations: a Gaussian $\Lambda$CDM reference model,
a $\chi^2$ model, a primordial void model ({\it PVM}) and 4 models with different levels of
quadratic non-Gaussianities in the primordial 
gravitational potentials ($Q$ models). We then describe
the setup of the non-Gaussian initial conditions.

\subsection[]{Common simulation parameters}
\label{sec:Models:General}

We employ the public versions of the $N$-body codes 
{\sc HYDRA} \citep{Cou95} and {\sc GADGET} \citep{SprGadget2001}
to simulate collisionless structure formation in the \lcdm 
cosmology presently favoured by the \emph{WMAP} observations
\citep{Spe03}, with parameters given 
in Table \ref{tab:FractionReconstructed}. When
constructing the initial conditions, we input primordial density field 
to the non-linear transformations when 
they are needed to generate the non-Gaussian field, and then smooth with 
the CDM transfer function of adiabatic fluctuations. 

The normalisation of the Gaussian case is set so that the value of the linearly
extrapolated $\sigma_{8}$ at $z=0$ is 0.9. This value provides only an 
approximate match to the
$z=0$ abundance of clusters with optically derived masses \citep{Bah03a}  which is then
overestimated, but it agrees with the \emph{WMAP} preferred values 
(\citealt{Spe03}, see also
\citealt{Teg03b} and references therein). We tune the normalisation of the non-Gaussian
models so that they produce the same abundance of massive clusters 
as the Gaussian simulation.
In the $\chi^2$ model (resp. {\it PVM}) we end up with a slight deficit 
(resp. excess) compared to the Gaussian case,
but the quadratic series of models agrees very well with the abundance
given by the Gaussian model.

The starting redshift for all simulations is $z=70$. 
The softening length 
is kept constant in comoving coordinates (set to 
one tenth of the mean interparticle separation). 
In what follows, we distinguish
for clarity between the ``initial'' (or ``primordial'') 
and ``unevolved'' density fields which correspond
to $z=70$ and $z=\infty$ respectively. Particles are distributed on a regular grid at
$z=\infty$, and then displaced with the usual scheme \citet{Zel70}
corresponding to the gravitational potential of the desired primordial density
field realized on a $128^3$ mesh ($64^3$ for the $Q$ models). 
In the {\it PVM} model, additional displacement is
imparted in regions covered by the primordial voids.

In the next paragraph, we review the main features of the 3 classes
of non-Gaussian models whose initial density fields we shall reconstruct
in Section~\ref{sec:ResultsNonGauss}.

\subsection[]{$\chi^2$ model}
\label{sec:Models:Isoc}

The $\chi^2$ model with one degree of freedom is 
phenomenologically attractive because it is strongly positively-skewed:
it could be a valid description of a possible scale-dependent 
non-Gaussianity present on cluster scales.

To realize the corresponding density field,
\begin{equation}
 \rho_{\rm init}({\bf x})=\varphi_{\rm init}^2({\bf x}),
\label{eq:Chi2Def}
\end{equation}
we square on the $128^3$ grid the scale-free $n_{s}=-2.4$
Gaussian random field $\varphi_{\rm init}$ to obtain a $n_{s}\sim-1.6$ $\chi^2$ density 
field $\rho_{\rm init}$. This slope
is slightly shallower than the theoretical value ($-1.8$) because 
of the incomplete mode coupling resulting from squaring
the field on the grid (see for example \citealt{WhiteM99,RB00}). 
We then apply the appropriate transfer function.
Moments of the one-point PDF of a $\chi^2$ density field with 
power-law power spectrum  (this holds on
sufficiently small scales given the isocurvature transfer function) 
are theoretically independent of the smoothing length
which is employed to compute them, once the shape of the smoothing window has been set
 (e.g. \citealt{Pee99a,WhiteM99}). In practice when one uses a computational
mesh, this independence does not hold because of the incomplete mode coupling.

Immediately after displacing particles from grid using the Zel'dovich
approximation, we measure higher-order moments of the resulting initial density
field in the simulation using a cloud-in-cell (hereafter CIC) first
order particle-to-grid assignment scheme, and  smooth the resulting field
with a top-hat kernel of radius 8\hMpcDot As expected, the density 
is strongly positively skewed:
we find $D_{3,8}=1.8$  and a kurtosis $D_{4,8}=9.5$. (The notations
and definitions are those of \citealt{Pee99a}: 
$D_{3,8}=\langle (\delta_{8}/\sigma_{8})^3\rangle$
and $D_{4,8}=\langle(\delta_{8}/\sigma_{8})^4\rangle\,-3\, 
\langle(\delta_{8}/\sigma_{8})^2\rangle^2$.)
These values have to be compared to the numerical values 
of 2.0 and 9.1 respectively found on
a $128^3$, 200 \hMpc grid realisation of  $\chi^2$ smoothed with an 8 \hMpc radius.
Here, we employ the $\chi^2$ case as a typical example
of models with a strongly non-Gaussian initial density field 
but still linear at the starting redshift
over all resolved scales (see \citealt{Mos91} for a similar simulation).

\subsection[]{Primordial voids model}
\label{sec:Models:PVM}

The physical motivation for primordial voids resides in a possible first-order
phase transition occurring in the early Universe in the framework of extended
inflation \citep{La89}. In some cases, incomplete nucleation 
of bubbles of true vacuum can result in the
late-time persistence of a network of voids of astrophysically relevant radii
which are completely empty of matter by the end of inflation/epoch of
reheating. The radii of the voids can then increase in comoving coordinates
once the Universe has entered the matter-dominated era (see \citealt{Bert85a} for the appropriate scalings).

Specifically, we employ for {\it PVM} the parameters used by \citet{Gri03}, 
and consider the voids as fully empty regions immediately surrounded by thin compensating shells of the
matter which were swept up during their expansion. At $z=0$, the cumulative number density of voids:
\begin{equation}
N_{V}(>r)=A\;r^{\alpha} \quad \rmn{with} \quad \alpha=-3
\label{eq:VoidRadDist}
\end{equation}
is integrated from $r_{\tx{min}}=10$ to $r_{\tx{max}}=25$ \hMpcDot The index $\alpha=-3$ is expected in viable
models of the first-order phase transition that we assume responsible for the primordial voids (see, e.g. \citealt{Occ94}).
The upper cutoff $r_{\tx{max}}$ approximately corresponds to the largest voids observed locally \citep{ElAd00,Pee01,Hoy03},
and $r_{\tx{min}}$ is a possible lower cutoff depending on details of 
the phase transition. The normalising constant $A$ is then
 set such that to obtain a 40 percent volume fraction of the voids in the present-day Universe.

We refer the readers to \citet{Ma03a} for the simple implementation of {\it PVM} in $N$-body simulations.
In short, we first displace the uniform grid of particles 
with the Zel'dovich prescription exactly
as it is done for the standard Gaussian models. Then, before the simulation is
started, we realize  the ``void network''.
We compute a list of voids with  centres randomly placed in the box and radii
set according to equation (\ref{eq:VoidRadDist}) (we ensure the voids are not 
initially overlapping). Next, we simply move
any particle which is located inside one of these voids to its edge, along a 
straight line passing over the void centre and this
particle's Lagrangian position. We assign the local shell 
velocity to the particle and then start the simulation.

On 8 \hMpc scales, the skewness and kurtosis of the 
high-redshift density field measured immediately
after constructing the initial conditions are $D_{3,8}=-1.68$ 
and $D_{4,8}=10.11$. In fact, the void+Gaussian
initial density field is both strongly negatively-skewed and 
non-linear on the scales of about $3$ \hMpcDot
As such, it constitutes an important testbed for the possible 
reconstruction of the density fields for which
shell crossing has already occurred at $z=70$.

\subsection[]{Quadratic contributions to primordial gravitational potentials}
\label{sec:Models:Quad}

The last four non-Gaussian cases which we study in this work, originate from
the same generic model. The primordial gravitational potential 
includes a quadratic (also called ``non-linear'')
contribution weighted by the parameter $f_{\rm{NL}}$ in addition to the
Gaussian field $\Phi_{\rm{init}}$ (see for example  \citealt{Ver01}):
\begin{equation}
\label{eq:case_B_def}
\Phi_{\rm{tot}}=\Phi_{\rm{init}}+
f_{\rm{NL}}\,(\Phi_{\rm{init}}^2-\langle \Phi_{\rm{init}}^2 \rangle).
\end{equation}
Equation (\ref{eq:case_B_def}) has become a convenient 
parametrisation of a small primordial non-Gaussianity
that would be expected in the simplest models of 
inflation \citep{Malda03}. (A non-Gaussianity written as a
quadratic correction to the primordial density field 
would be closer to a model with topological defects.)
It has been found that higher-order correlation functions of the CMB temperature maps
are significantly more efficient than those of LSS to constrain a 
non-Gaussianity such as that given by equation (\ref{eq:case_B_def}) 
\citep{Ve00,Ver01}. The value of $f_{\rm{NL}}$ is currently constrained 
to lie between -58 and 134 \citep{Kom03}. Here, we simulate models
with $f_{\rm{NL}}$ of $\pm50$ and $\pm100$ ($f_{\rm{NL}}=-100$
is used to have a symmetric set). Positive
(resp. negative) values of $f_{\rm{NL}}$ result in a deficit
(resp. an excess) of the abundance of massive clusters at $z=0$ compared
to the Gaussian, $f_{\rm{NL}}=0$, case.  Note that definition (\ref{eq:case_B_def}) implies
that the level of non-Gaussianity in $\Phi_{\rm{tot}}$ depends on
both $f_{\rm{NL}}$ and  $\langle \Phi_{\rm{init}}^2 \rangle$. Some 
caution is therefore necessary when applying the
nonlinear transformation to $\Phi_{\rm{init}}$ on the initial 
condition grid so that the non-Gaussianity effectively
realized in the simulation on LSS scales corresponds to that 
probed in the CMB maps with the same
$f_{\rm{NL}}$. This is done in the following four steps. First, we 
realize a Gaussian random potential
field $\Phi_{\rm{init}}$ with $n_{s}=-3$ (corresponding to a density 
field with $n_{s}=-1$). Second, we normalise it so
that its extrapolation to the $l\sim10$ scales probed 
by \emph{COBE} \citep{Bun96} with conversion of
the resulting rms to CMB temperature fluctuations assuming adiabatic 
modes corresponds to the value
measured by \emph{COBE} \citep{Wright94}. Third, we realize 
equation (\ref{eq:case_B_def}) using the normalised
$\Phi_{\rm{init}}$. Finally, we transform back $\Phi_{\rm{tot}}$ to a 
density field and normalise
it to the adequate value for the matter 
power spectrum normalization, $\sigma_8$, at the starting 
redshift. Thereafter, the setup of the initial conditions
is the same as that for the Gaussian simulation.

Compared to the {\it PVM} and $\chi^2$ cases, the quadratic models 
considered here fall within CMB
constraints and are significantly closer to Gaussian models.
As a result, we expect that distinguishing between these models 
and a Gaussian primordial density field
by reconstruction techniques to be much more difficult for the $Q$
models than for the $\chi^2$ or the {\it PVM} models.

Next, we present the main aspects of the method employed to reconstruct the initial
displacement (and density) field from the $z=0$ outputs of 
our Gaussian and the 6 non-Gaussian models.



\section{Reconstruction of density fields at high redshifts}
\label{sec:Method}

In this section,
we start by reviewing the main features of the MAK reconstruction
method. Then, we detail the various particle samples which we have used 
for the reconstructions.

\subsection{Overview of the algorithm}
\label{sec:Method:Algo}

On sub-\hMpc to \hMpc scales, significant multistreaming takes place 
in the low-redshift Universe, so that
the pressureless Eulerian fluid approach to structure formation fails \citep{Pee87}.
Reconstruction is a well-posed problem insofar as significant multistreaming has not occurred.
Our main hypotheses are that the mapping ${\bf q}\rightarrow{\bf x}$ between the
Lagrangian coordinate, ${\bf q}$ at $t_{\rm init}$, and the Eulerian
coordinate of a particle, ${\bf x}({\bf q},t_{\rm final})$, can
be written as the {\it gradient} of a {\it convex} potential $\Phi({\bf q},t)$. The 
convexity guarantees that the mapping is one-to-one,
and therefore the existence of the reciprocal map ${\bf x}\rightarrow{\bf q}$. In
addition, these hypotheses imply that reciprocal 
mapping also derives from a convex potential
 $\Theta ({\bf x},t)$, and that $\Theta$ and $\Phi$ are related by 
the Legendre-Fenchel transform \citep{Arn78}.
The goal of the algorithm is to obtain the mapping $\Theta$. This is 
addressed in the first paragraph that follows.
The details of the computation of the reconstructed peculiar velocities 
is discussed in the second paragraph that follows.

\subsubsection{From cosmological reconstruction to the assignment problem}
\label{sec:Method:Algo:Euler2Lagrange}

Substituting the inverse Lagrangian map, ${\bf q}=\bf\nabla_x \Theta({\bf x})$,
in the equation of conservation of mass,
$\rho({\bf x})\rmn{d}{\bf x}= \bar \rho\; \rmn{d}{\bf q}$ yields 
the so-called Monge-Amp\`ere (MA) equation:
\begin{equation}
{\rm det}\left[{\partial^2 \Theta({\bf x})\over \partial x_i \partial x_j}\right]=
{\rho({\bf x})\over \bar\rho}\;,
\label{eq:MongeAmpere}
\end{equation}
where $\bar \rho$ is the unevolved (uniform) Lagrangian density and  $\rho({\bf x})$ the evolved
Eulerian density. The linearisation of equation (\ref{eq:MongeAmpere}) yields 
the Poisson equation. Recently,
it has been shown that the map that is solution to the MA equation is the
unique solution to the Monge-Kantorovich optimisation problem (hence ``MAK''),
in which one seeks the map
${\bf x}\rightarrow {\bf q}$ which minimises the quadratic cost function:
\begin{equation}
I=\int_{\bf q} \bar\rho\vert{\bf x}-{\bf q}\vert^2 \rmn{d}^3q=
\int_{\bf x} \rho({\bf x})\vert{\bf x}-{\bf q}\vert^2 \rmn{d}^3x
\label{eq:cost}
\end{equation}
(see \citealt{Bena00}, FMB, and references therein for 
the detailed proofs). In practice, when the method is
applied to discrete points such as galaxies or $N$-body particles at $z=0$, one discretizes the cost
function~(\ref{eq:cost}) into:
\begin{equation}
I=\min_{j(i)}\left(\sum_{i=1}^N\left({\bf q}_{j(i)}-{\bf x}_i\right)^2\right)\;,
\label{eq:assign}
\end{equation}
which is known as the assignment problem. We stress here for clarity
that both the set of initial and 
final positions at $z_{\rm init}$ and $z_{\rm final}$ are known, but the
correspondence between the two is not. The set of final
positions is measured from the observations or the simulations,
that of the initial positions is produced using the fact that the
unevolved distribution is uniform: in our case, it corresponds 
to the nodes of a regular mesh. Given $N$ initial and $N$ final
entries, the aim is to  find the permutation which minimises
the quadratic cost function \footnote{Because the simulation
boxes are periodic, optimal assignment is found
after taking his periodicity into account.}.

The simplest algorithm which would solve the 
assignment problem (\ref{eq:assign}) would clearly
have a factorial complexity: one needs
to search among $N!$ possible permutations for the one which has the
minimum cost. However, advanced assignment algorithms exist which
reduce the complexity of the problem from factorial to polynomial. The latest
algorithm developed by M. H\'enon and used in this work, which is a cosmological adaptation of
Bertsekas' auction algorithm, scales approximately
as $N^{2.5}$ (for relevant 
details see for example \citealt{Hen92,Hen95,Bertse98}).

\subsubsection{Reconstruction of the peculiar velocity field}
\label{sec:Method:Algo:PecVels}

Once we have obtained the optimal one-to-one mapping between the present-day
and its Lagrangian position for every particle, we can evaluate 
the present, reconstructed peculiar velocities for
these particles using the Zel'dovich approximation. From there, their 
reconstructed positions can be 
interpolated at any desired redshift
between $z_{\rm init}$ and $z_{\rm final}$. This works as follows.
In the Zel'dovich approximation, peculiar velocities are given by :
\begin{equation}
\dot{\bf x}=f(\Omega)\,H(t)\,({\bf x}-{\bf q})
\label{eq:Zeldovich}
\end{equation}
where $f(\Omega)=\rmn{d}{\rm ln} D/\rmn{d}{\rm ln} a$ is the dimensionless 
linear growth rate, $D(t)$
is the amplitude of the growing mode today, $a$ is 
the cosmic scale factor and $H(t)$ is the value of the
Hubble parameter. We note $D_0$ and ${\bf x}_0$ the growth
factor and the known present-day position of 
a particle respectively. If we now assume
the validity of the Zel'dovich approximation throughout, we can
integrate equation (\ref{eq:Zeldovich}) backwards from today (recall
that ${\bf q}$ is known from the MAK mapping) and obtain the position at
any $z$ using:
 \begin{equation}
{\bf x}(z)={\bf q}+{D(z)\over D_0}\left({\bf x}_0-{\bf q}\right)
\label{eq:PosScaling}
\end{equation}
Equations~(\ref{eq:Zeldovich}) and~(\ref{eq:PosScaling}) can 
be regarded as a ``backwards'' Zel'dovich
scheme going from $z=0$ to high redshifts to differentiate 
it clearly from the ``forward'' Zel'dovich scheme which is employed
to go from $z=\infty$ to lower redshifts ({\it e.g.} $z=70$ in our case in
order to set the initial conditions of the simulations).  As we look for the map
satisfying the MA equation in the following sections, we 
start from the $z_{\rm final}=0$ distribution
of particles in the simulations and relate it to an unevolved position
which is a node of a regular lattice. Before we turn 
to the detailed analysis of the reconstructed high-redshift fields, we
define in the next paragraph the series of particle samples 
that we have used for our reconstructions.

\subsection{Defining particle samples for reconstruction}
\label{sec:Method:PartSample}

For each simulation, we have traced back a set  $S_{1}$ of $64^3$ particles. These particles
are regularly selected at $z=\infty$ from the uniform $128^3$ grid of particles, then
located at $z=0$ in the full, $128^3$ simulation output, and 
finally used for the reconstruction. In the case of
the $Q$ series of models with $64^3$ particles, there is no resampling needed. The
mean interparticle distance in $S_{1}$ is 3.1 \hMpcDot One motivation for this was
to reduce the computational cost of our $7$ reconstructions.

In the Gaussian and the $\chi^2$ cases, we have also reconstructed the full set $S_{0}$
of $128^3$ particles, to help our theoretical interpretation of MAK 
and to verify that our results are not affected
by the under-sampling performed in $S_{1}$. Finally, for the Gaussian, the
$\chi^2$ and the {\it PVM} models, we
have reconstructed a smaller, $64^3$ particle ``sub-box''  extracted from the 200 \hMpc box,
we call it ``dense samples'' $S_{2}$; it is obtained as follows. First, we
have extracted a 100  \hMpc-size square box of $64^3$ particles
from the original Lagrangian grid, and located them in the final simulation
output. Then, we have reconstructed their final positions back in time.
We will use this sample together with the sample $S_{0}$ as we discuss issues
of resolution.  The summary of the number of particles, boxsize and mean interparticle
separation $\overline{l}$ of our various particle samples is given 
in Table~\ref{tab:FractionReconstructed}.

\begin{table}
\caption{Parameters of the various particle samples (full box, sparse samples and dense sample, 
respectively $S_{0}$, $S_{1}$ and $S_{2}$) of our 7 models used 
to test reconstruction. Note that although we have 
called ``full box''  the particle samples
of the $Q$ models which only have $64^3$ particles in total, their mean 
interparticle separation, $\overline{l}$, makes them more similar to the  ``sparse sample'' cases.
As a result we also call them $S_{1}$. Also given is the fraction 
of particles that the MAK reconstruction scheme reassigns to
their true Lagrangian position on the uniform grid. We call 
these particles ``ideally reconstructed''.
}
\begin{tabular}{@{}lllll@{}}
\hline
Sample  &   N & Box & $\overline{l}$  &  ideal \\
           &    & [ \hMpc ] & [ \hMpc ]  &   \\

\hline\hline
\multicolumn{5}{|c|}{Gaussian}\\
\multicolumn{5}{|l|}{$\, \Omega_m=0.3\,,
\,h=0.65\,,\, \sigma_8=0.99\,,\,n_s=1\,,\, M=3.2\times10^{11} h^{-1} M_\odot$}  \\
\hline
full box ($S_{0}$)  & $128^3$ & 200 & 1.5 & 18\%   \\
sparse sample ($S_{1}$)  & $64^3$ & 200 & 3 & 37\%  \\
dense sample ($S_{2}$)  & $64^3$ & 100 & 1.5 & 17\%   \\
\hline\hline
\multicolumn{5}{|c|}{$\chi^2$}\\
\multicolumn{5}{|l|}{$\, \Omega_m=0.2\,,
\,h=0.7\,,\, \sigma_8=0.61\,,\,n_s=-2.4\,,\, M=2.1\times10^{11} h^{-1} M_\odot$}  \\
\hline
full box ($S_{0}$)  & $128^3$ & 200 & 1.5 & 31\%   \\
sparse sample  ($S_{1}$)  & $64^3$ & 200  & 3 & 60\%   \\
dense sample ($S_{2}$)  & $64^3$ & 100 & 1.5 & 33\%   \\
\hline\hline
\multicolumn{5}{|c|}{{\it PVM} }\\
\multicolumn{5}{|l|}{$\, \Omega_m=0.3\,,
\,h=0.7\,,\, \sigma_8=0.9\,,\,n_s=1\,,\, M=3.2\times10^{11} h^{-1} M_\odot$}  \\
\hline
sparse sample ($S_{1}$)  & $64^3$ & 200   & 3 & 28\%    \\
dense sample ($S_{2}$)  & $64^3$ & 100 & 1.5 & 10\%   \\
\hline\hline
\multicolumn{5}{|c|}{$Q$ models}\\
\multicolumn{5}{|l|}{$\, \Omega_m=0.2\,,
\,h=0.7\,,\, \sigma_8=0.9\,,\,n_s=1\,,\, M=2.5\times10^{11} h^{-1} M_\odot$}  \\
\hline
\multicolumn{5}{|c|}{$Q_{-100}$}  \\
\hline
full box ($S_{1}$)  & $64^3$ & 200  & 3 & 43\%   \\
\hline\hline
\multicolumn{5}{|c|}{$Q_{-50}$}  \\
\hline
full box  ($S_{1}$)   & $64^3$ & 200 & 3 & 43\%   \\
\hline\hline
\multicolumn{5}{|c|}{$Q_{+50}$}  \\
\hline
full box  ($S_{1}$)  & $64^3$ & 200 & 3 & 42\%   \\
\hline\hline
\multicolumn{5}{|c|}{$Q_{+100}$}  \\
\hline
full box  ($S_{1}$)   & $64^3$ & 200 & 3 & 41\%   \\
\hline
\end{tabular}
\label{tab:FractionReconstructed}
\end{table}

The reconstruction of the $128^3$ particles of the Gaussian 
and $\chi^2$ sample $S_{0}$ is unprecedented and
breaks a computational barrier for cosmological 
reconstruction schemes. 

\begin{figure*}
\centerline{
\includegraphics[width=8cm]{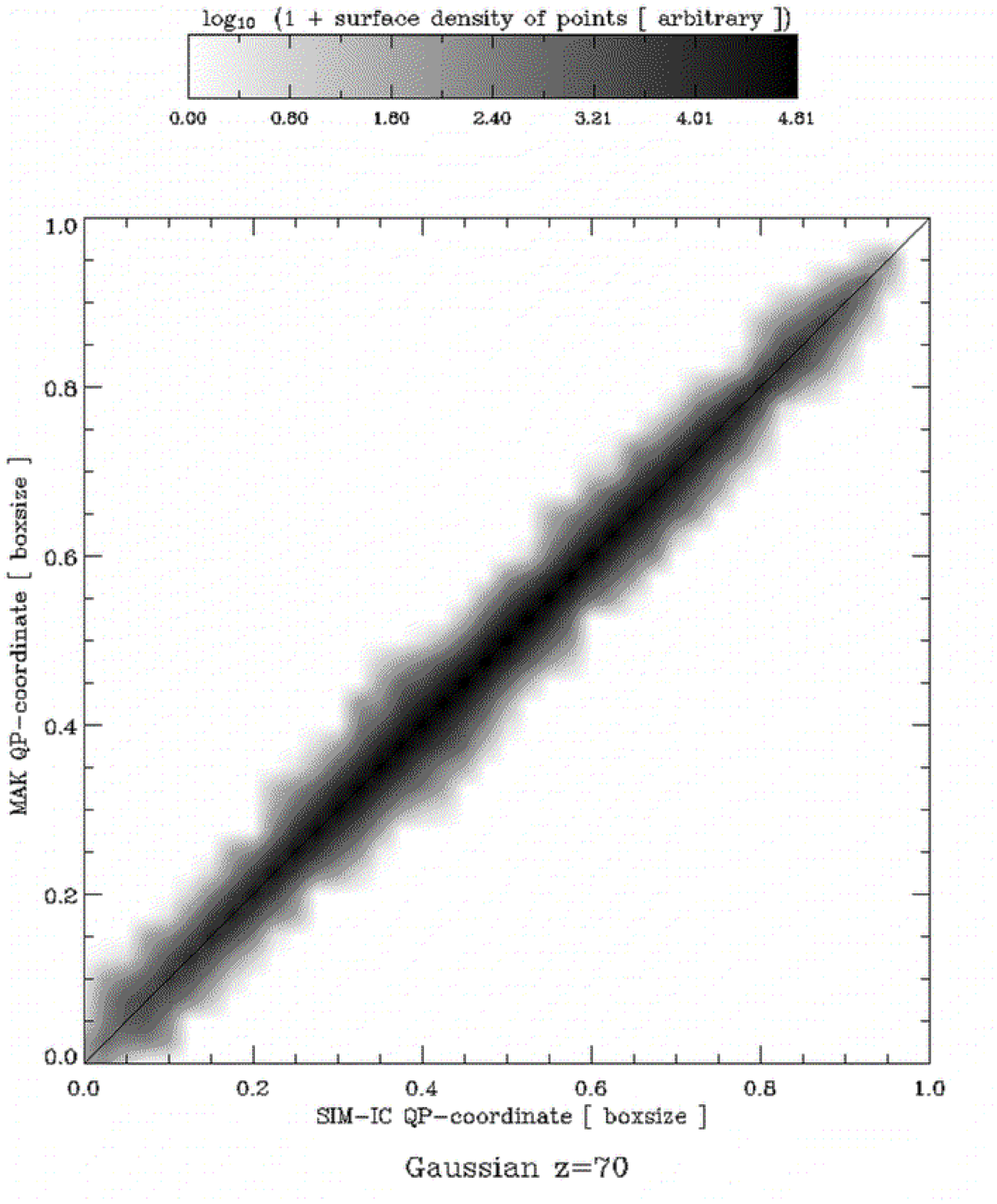}\includegraphics[width=7.2cm]{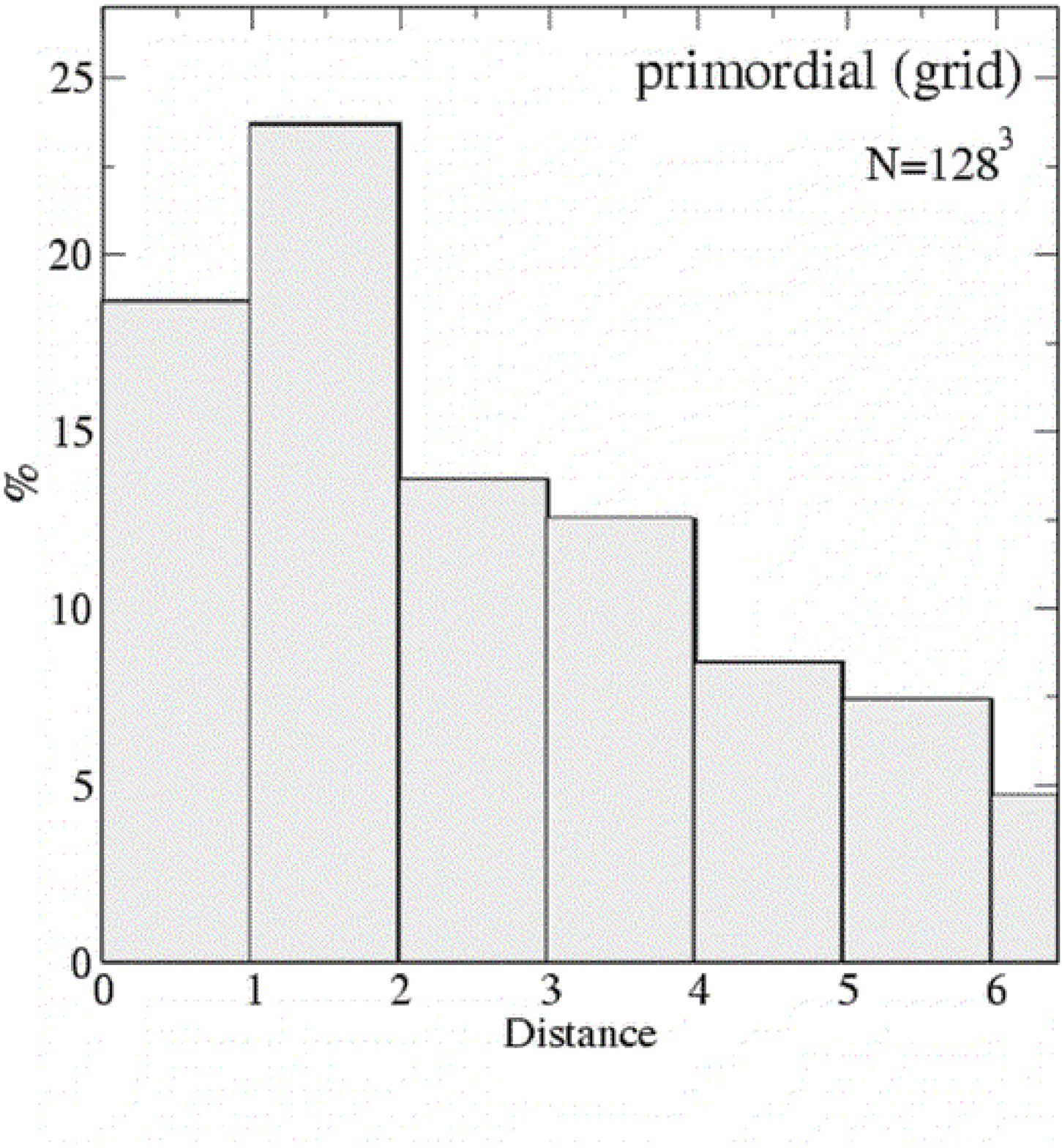}
}
\caption{
Performances of MAK reconstruction.
{\it left panel:} Scatter plot of the MAK-reconstructed 
initial coordinates of a particle versus
its  initial, post-Zel'dovich coordinate for 
the $128^3$ particles ($S_0$ sample) in the Gaussian simulation.
We use a ``quasi-periodic projection'' (QP) coordinate 
$\tilde q=(q_1+q_2\sqrt 2+q_3\sqrt 3)/(1+\sqrt 2+\sqrt 3)$,
with $q_1$, etc... $\in [0,1]$ where 1 corresponds to 
the box size. These QP coordinates guarantee a 1-to-1 correspondence between
the $\tilde q$-values and the points on a regular grid. To clearly show 
where the majority of particles reside on the
diagram, we show the decimal logarithm of 1 plus the local density of 
points (the resolution is 1/256 in QP coordinate,
and the Lagrangian mesh spacing is 1/128). Note that most of the particles lie
almost perfectly on the diagonal already illustrating the success of MAK.
{\it right panel:} Histogram of the distribution of the distances, expressed
in units of mesh size, between the
true Lagrangian, discrete, positions of each particle on the uniform grid
and the position assigned to it by the MAK reconstruction. The first bin
corresponds to what we call "ideal" reconstruction, as shown in Table 1. If
one adds up the percentages obtained in the first and the second bin, one
should obtain a good estimate of the percentage of "ideally" reconstructed
positions for the sparse sample $S_1$, as is the case (see Table 1).
}
\label{fig:Scatter_128Gauss}
\end{figure*}

In Fig.~\ref{fig:Scatter_128Gauss}, we show for this simulation
the scatter plot of the MAK-reconstructed Lagrangian positions of 
particles against their true Lagrangian positions, as well as the histogram of the
differences between true and reconstructed 
positions. This very simple preliminary analysis already shows the success of
MAK reconstruction, at least on large scales. However, we see that "exact" reconstruction is achieved
for only a small fraction of particles. The collapse of highly nonlinear
structures indeed brings about shell crossings and local mixing.
MAK is therefore unable to distinguish between particles belonging to the same  
condensed object as will be illustrated in detail later.
In fact, we shall see that this does not pose an obstacle to the recovery of 
the primordial density field. 

\section{Efficiency of MAK as a nonlinear reconstructor}
\label{sec:ResultsGauss}

This section focuses on the Gaussian model.
Firstly, we present detailed comparisons between the 
MAK-reconstructed displacement field with the non-linear displacement field (here
after SIM NL) which it is designed to 
recover. Secondly,  we compare the reconstructed and simulated nonlinear 
``density fields'' that will be given by minus
the divergence of the displacement field. We denote the latter by 
${\bf v}$ and for this reason on occasions we refer to it as
the ``velocity field''. In both cases, comparisons 
with the Zel'dovich, initial
conditions fields (hereafter SIM IC) are also given.  Thirdly, we compare the
present velocities found by MAK to the true peculiar velocities given 
by the simulation: the agreement on large scales is
a notable achievement of our reconstruction method.
In the last paragraph, we discuss potential resolution issues.

Except for Section \ref{sec:ResultsGauss:PecVels} which 
concerns the present peculiar velocity field,
the analyses are performed on the displacement field which is a Lagrangian
quantity, always estimated as a function of Lagrangian coordinate, ${\bf q}$,
on the primordial grid of initial unperturbed  positions of the particles.
In all cases, the measurements use the actual displacement between
$z=\infty$ (the unperturbed grid) and $z=0$. For SIM IC, this requires a scaling
of the linear displacement field using the Zel'dovich approximation.

Note furthermore that all the analyses in this section and in
forthcoming ones are performed with additional top hat smoothing applied
to the fields with spherical windows of various radii ($3\ h^{-1}$, $8\ h^{-1}$ Mpc, 
and sometimes $12\ h^{-1}$ Mpc). It is important to emphasize that
top hat smoothing with a spherical window of radius
$R$ is approximately equivalent to Gaussian smoothing with a window of
radius $R/\sqrt{5}$. This should be kept in mind when comparing with
previous works which usually use Gaussian smoothing
({\it e.g.} Monaco et al.~2000).


\subsection{Comparison of the displacement fields}
\label{sec:ResultsGauss:Displ}

In the present and the next paragraph we analyse the $S_{1}$ sample of the 
Gaussian model to show that, as expected from
the description of the algorithm in Section~\ref{sec:Method},  MAK is an 
excellent tracer of the non-linear displacement field. 
We recall that the SIM NL displacement 
is simply measured in the simulation by joining a particle's
position on the Lagrangian grid to that particle's position 
in the $z=0$ simulation output.  

\begin{figure*}
\includegraphics[width=16cm]{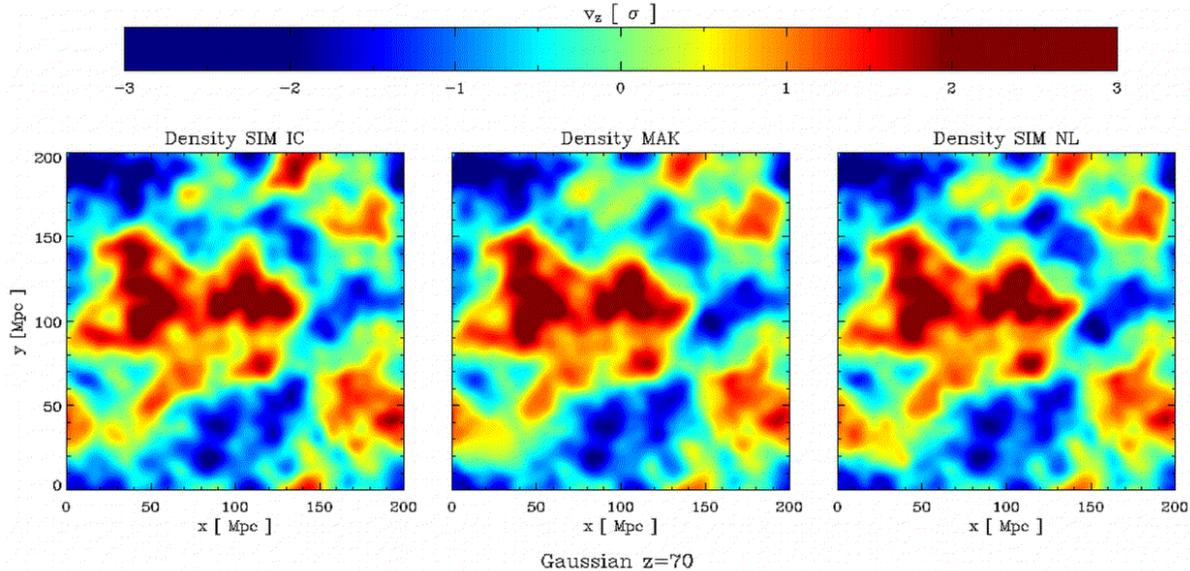}
\caption{Projected $v_{\rmn{z}}$  displacement field taken 
from a 20 \hMpc-thick slice normal to the z-direction,
for the Gaussian model. The resolution before projection is $\sim$ 3 \hMpc 
in each direction. The left
panel is the Zel'dovich initial conditions (SIM IC),  the central panel 
is the MAK reconstruction and the right
panel is the nonlinear displacement SIM NL. (Note that we will keep this tryptich 
structure and the corresponding SIM IC, 
MAK and SIM NL association for all remaining colour images.) The colour scale 
shown at the top is common to all plots 
and expressed in multiples of the rms of  $v_{\rmn{z}}$ of each field. The 
largest and broadest features repeat 
well in all three maps, but the MAK and SIM NL maps also match in detail.}
\label{fig:velz_z70_Gauss}
\end{figure*}

\begin{figure*}
\begin{tabular}{@{}c@{}}
\includegraphics[width=8cm]{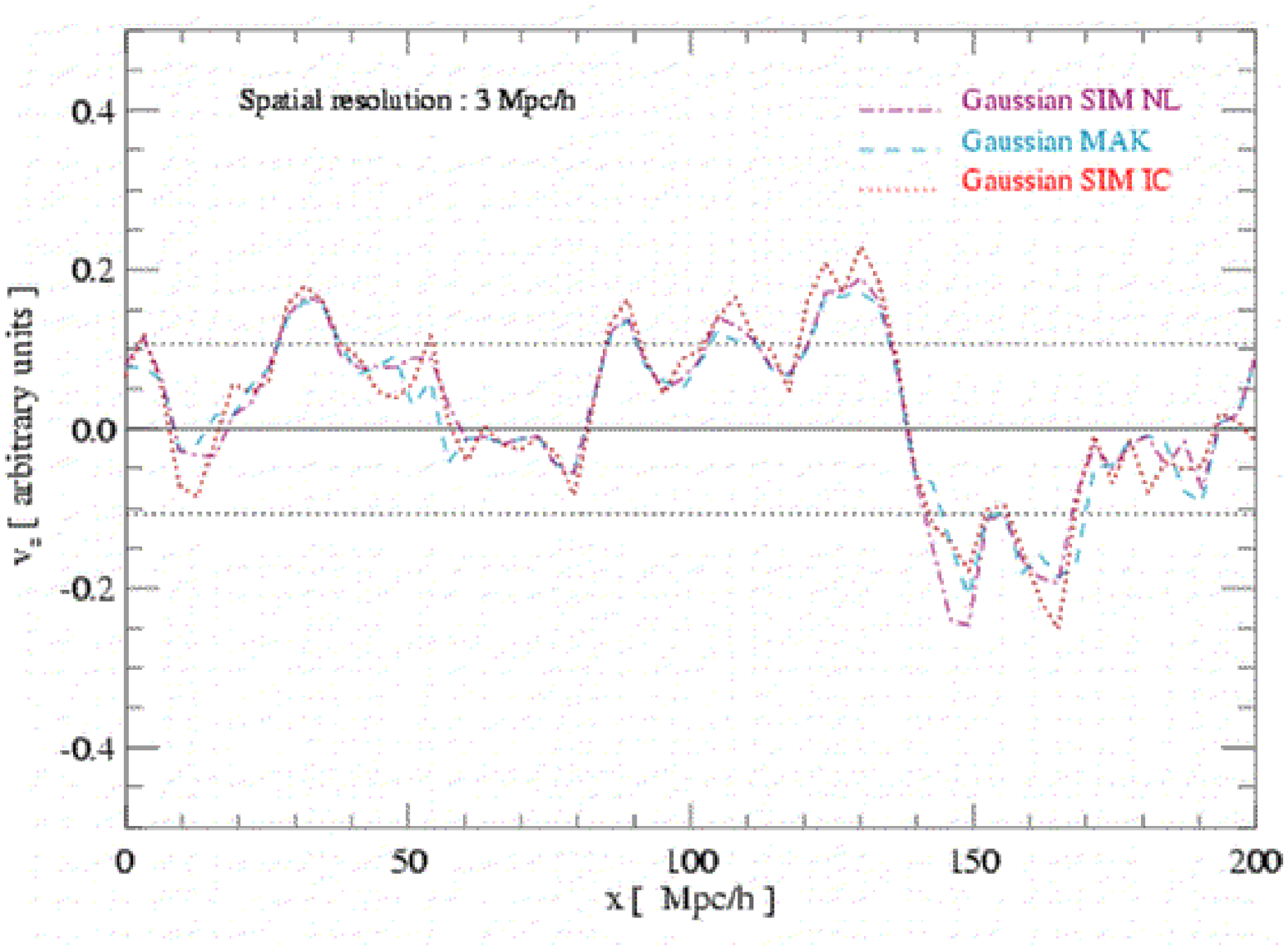} \\
\includegraphics[width=8cm]{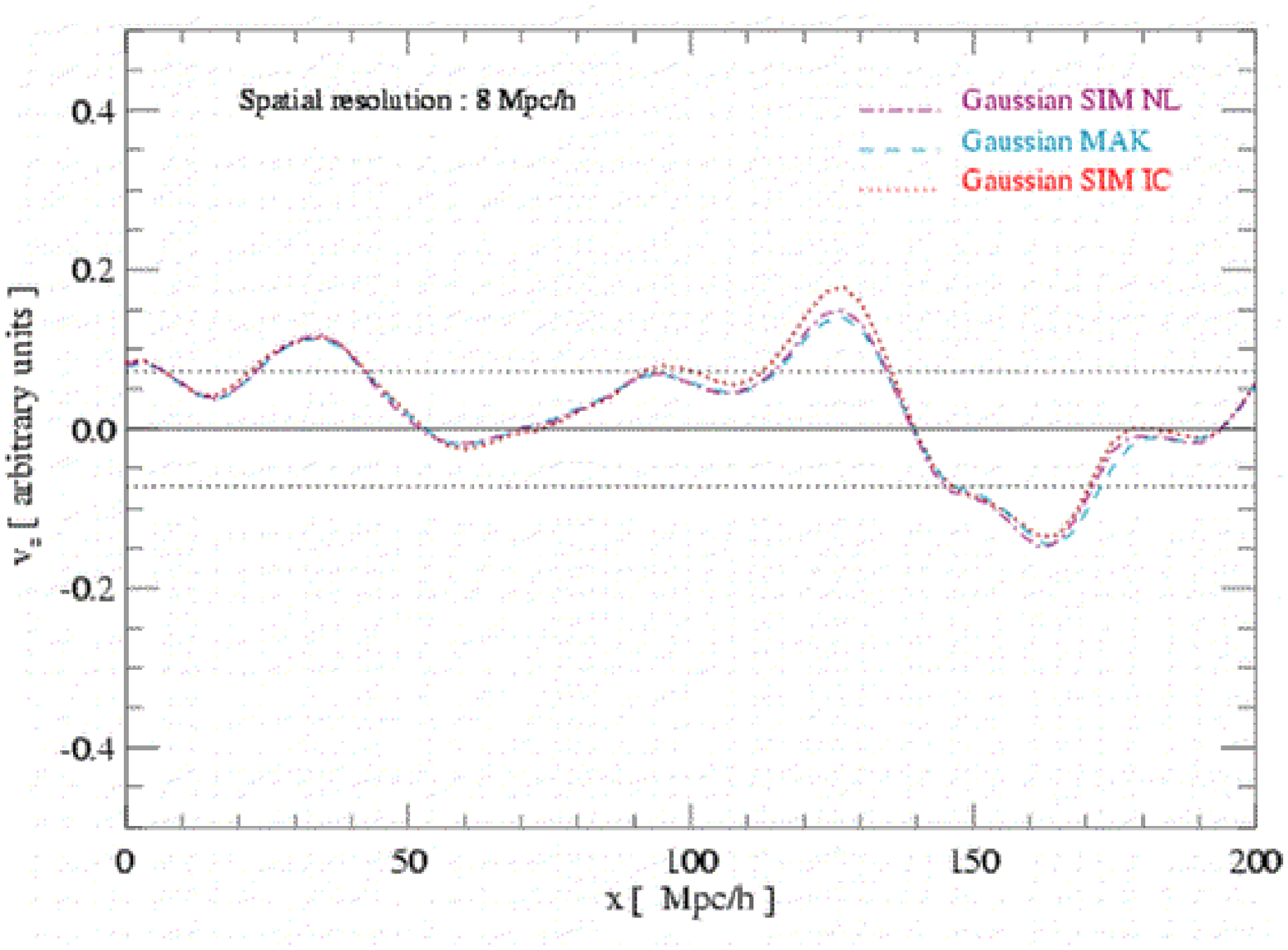}
\end{tabular}
\caption{Profiles of the initial z-component of the displacement fields 
$v_{\rmn{z}}$ in the Gaussian model, measured
 along a line parallel to the x-direction and located in the 
middle of the box. The resolution is 3 (resp. 8) \hMpc in the upper
(resp. lower panel). The Zel'dovich SIM IC, MAK reconstruction 
and SIM NL profiles are shown with dotted, dashed and dash-dotted
lines respectively. The horizontal dotted lines give $\pm \sigma$ of 
the SIM IC variations (the rms is measured from the profiles).
Note how well MAK reproduces the SIM NL displacement field, even at 3 \hMpc resolution.}
\label{fig:TestLine_vz_Smooth0_Gauss}
\end{figure*}

Fig.~\ref{fig:velz_z70_Gauss} compares, for SIM IC, MAK and
SIM NL, the z-components  of the displacement field projected along the
z axis from a 20 \hMpc thick slice.  The measurements have been performed
using a nearest grid point (NGP) interpolation of the particles to a $64^3$
grid, followed by smoothing with a top-hat of radius 
3 \hMpc.
There is a very good agreement between the three panels of the figure.
To have additional insight,
Figure \ref{fig:TestLine_vz_Smooth0_Gauss}, 
displays $v_{\rmn{z}}$ measured along a line. 
The two panels correspond to smoothing scales 3 \hMpc
and 8 \hMpc. This figure highlights the subtle differences between
the three panels of Fig.~\ref{fig:velz_z70_Gauss}. In particular we
see that the agreement between SIM NL and MAK is better than between
SIM NL and SIM IC or between MAK and SIM IC. Note that even at 
mildly nonlinear scales, 3 \hMpcCom the agreement between MAK and SIM NL
is surprisingly good.

\begin{figure*}
\includegraphics[width=8cm]{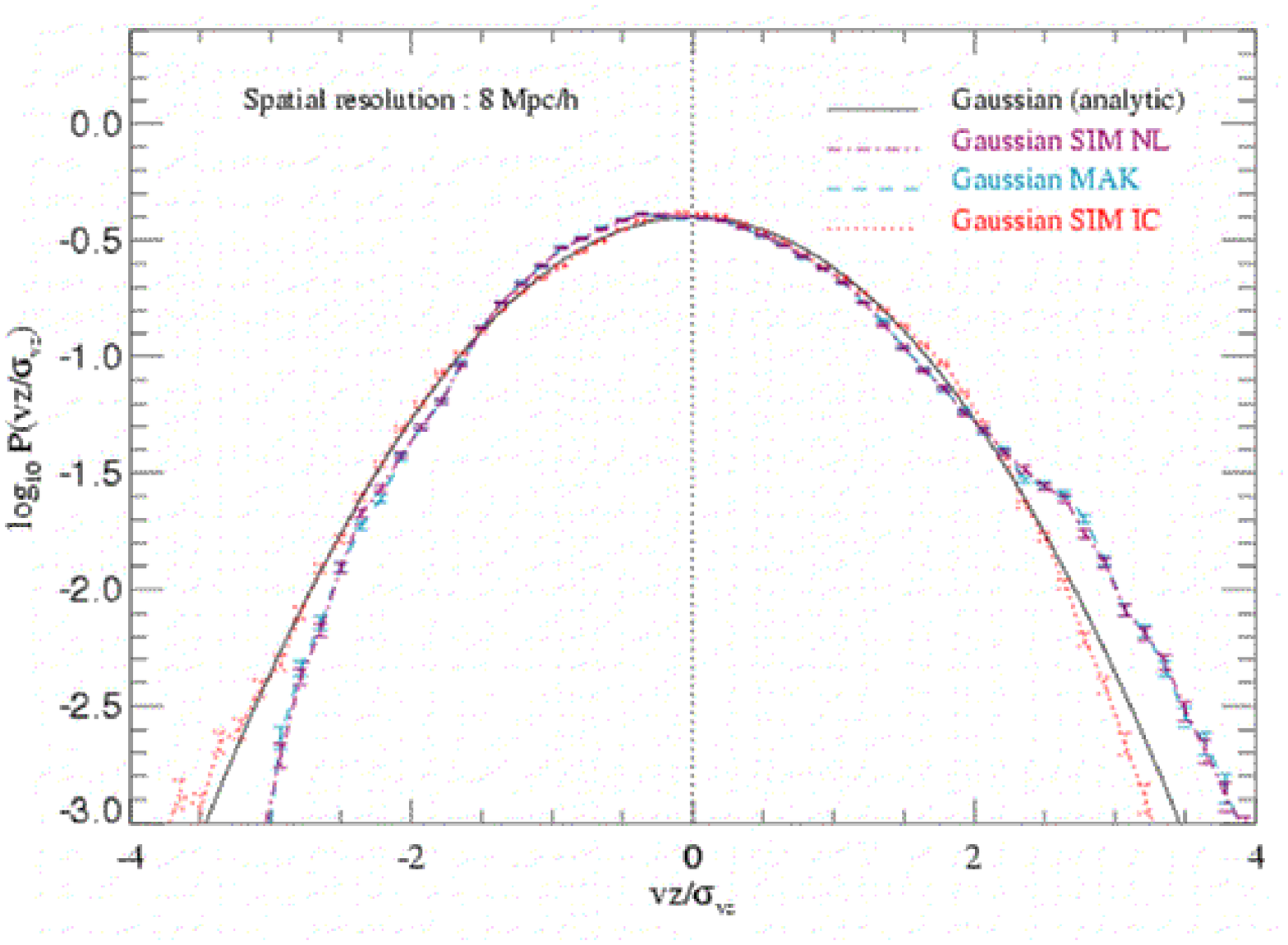}
\caption{PDFs of the $v_{\rmn{z}}$ displacement for the
SIM IC (dotted line) and MAK reconstruction (dashed line)  of 
the Gaussian model (rescaled to unit rms).
The fields have been rescaled to unit rms.}
\label{fig:velz_PDF_Gauss}
\end{figure*}

These results are confirmed by Fig.~\ref{fig:velz_PDF_Gauss} which examines
the PDFs of the $v_{\rmn{z}}$ component of the displacement field, for SIM NL,
MAK and SIM IC, at 3 \hMpc: the curves corresponding to MAK and SIM NL are nearly
superimposed. They present a slight skewness which is due in the SIM NL case
to non-Gaussianity brought by gravitational clustering. We already see that
the latter feature is entirely captured by MAK.

So far we have compared single components of the displacement field: we now 
address the differences and/or similarities in amplitude
and in direction between the three displacement fields as a function of 
the underlying over-density. 

\begin{figure*}
\includegraphics[width=16cm]{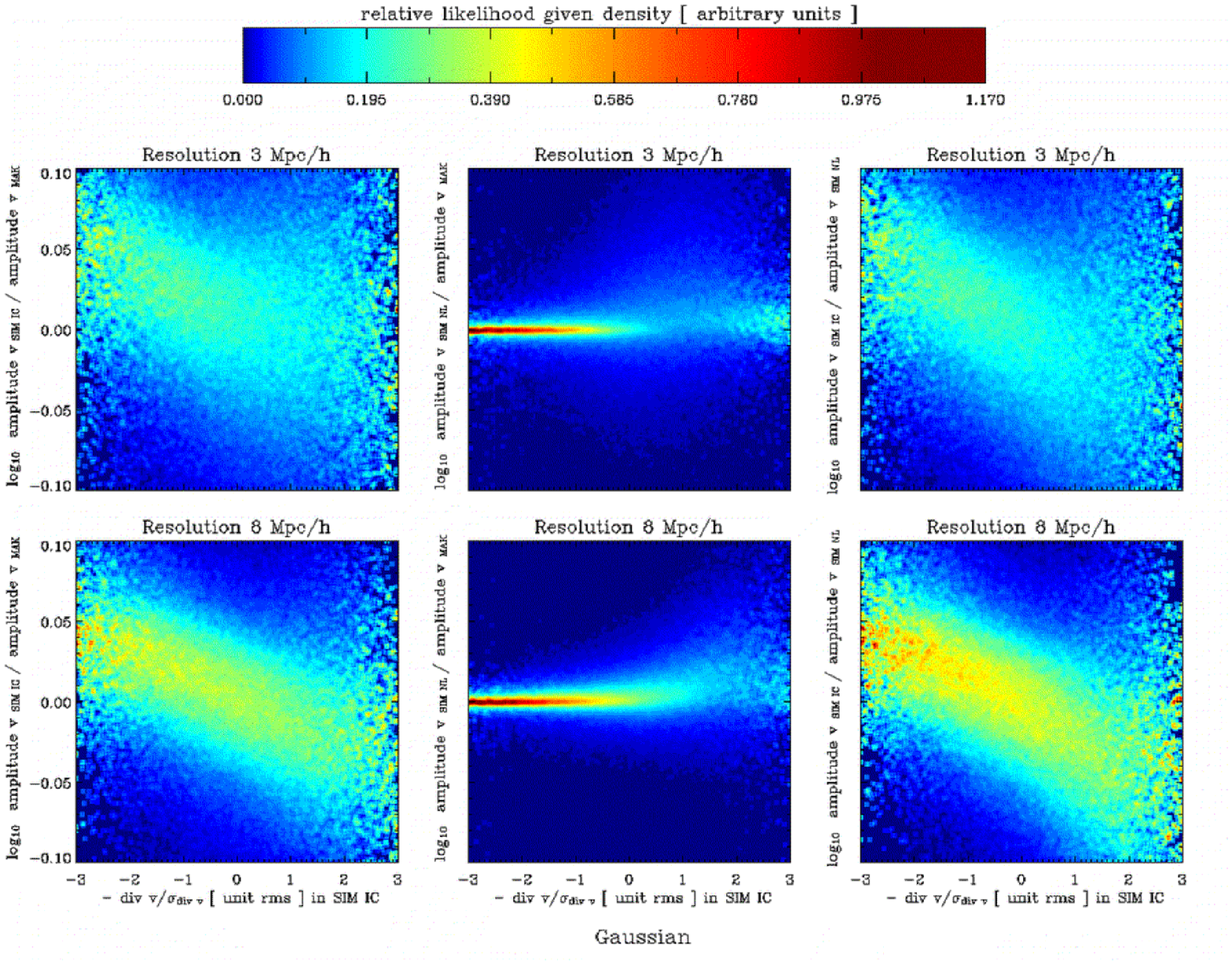}
\caption{Left panels show the conditional PDF of the local ratio between
the amplitudes of the Zel'dovich SIM IC and of the MAK-reconstructed
displacements, as a function of the 3 \hMpc local over-density in
the initial conditions SIM IC. The 
plot corresponds to the Gaussian model. The top (resp. bottom) row is 
for a 3 (resp. 8) \hMpc smoothing
length. Middle panels: same
as the left panels, but for the ratio between the amplitudes
of the non-linear displacement field SIM NL and the reconstructed
displacement field. Right panels: same as the left panels, but 
showing the ratio between the amplitudes
of the Zel'dovich, SIM IC displacement field and the non-linear displacement field SIM NL.}
\label{fig:velAmplitudes_Gauss}
\end{figure*}

Figure \ref{fig:velAmplitudes_Gauss} displays the measured distribution 
of the ratios $v_{\rm SIM\, IC}/v_{\rm MAK}$, $v_{\rm SIM\, NL}/v_{\rm MAK}$ and $v_{\rm SIM\, IC}/v_{\rm
  SIM\, NL}$ as a function of initial density estimated as minus the divergence
of the initial displacement field. 
The upper and lower rows of panels correspond to a smoothing scale of 3 and 8
  \hMpc respectively, prior to the calculation of
  their ratios. 
Again, the amplitudes of the MAK and SIM NL displacements 
agree well. The match is slightly less good for initially over-dense
regions, as expected. Indeed, these regions might have experienced shell-crossing
during dynamical evolution.  Since SIM NL and MAK agree rather well, we notice
similar discrepancy between SIM NL versus SIM IC and MAK versus SIM IC.

\begin{figure*}
\includegraphics[width=16cm]{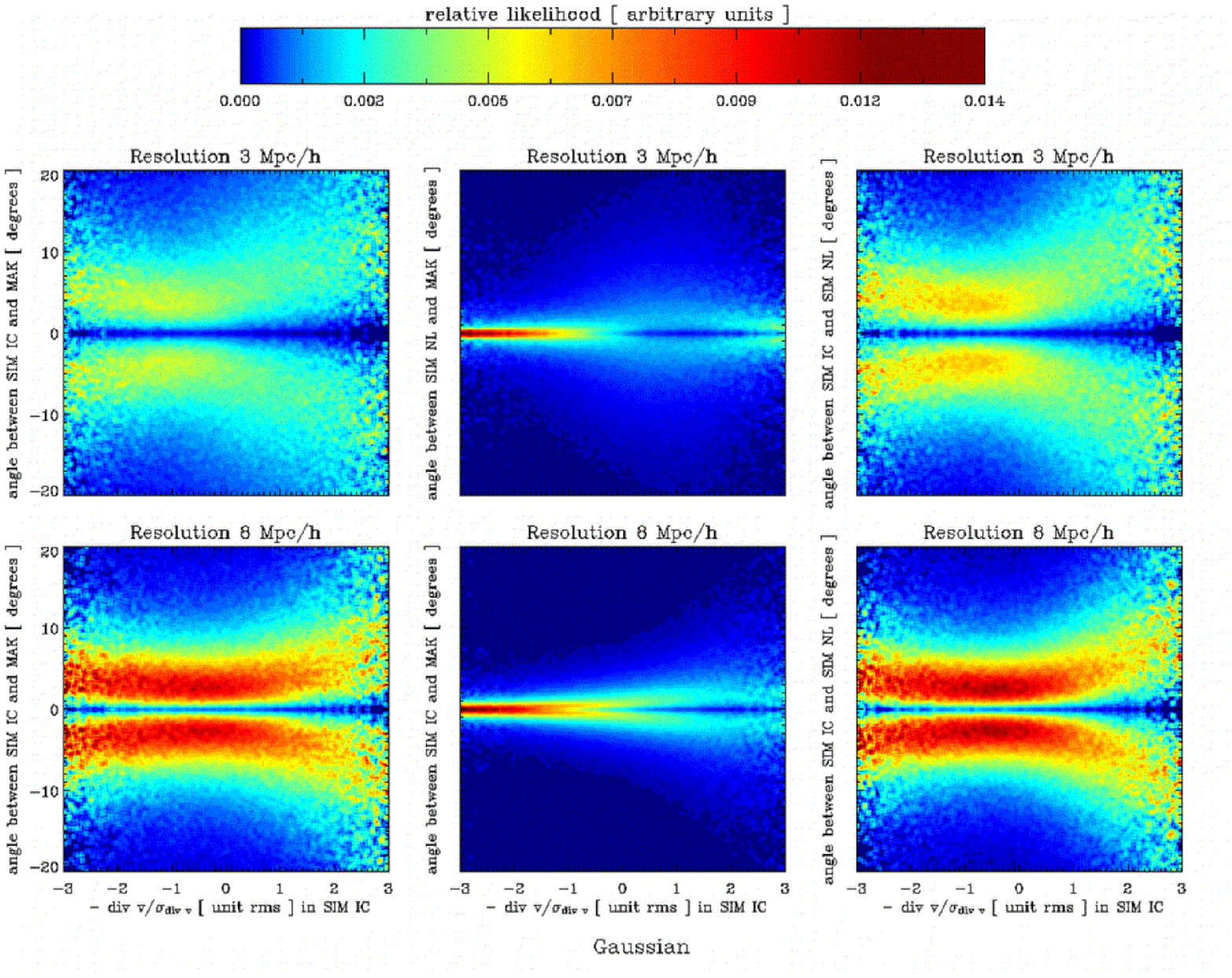}
\caption{Same as Fig.~\ref{fig:velAmplitudes_Gauss} but for the angle [in degrees] 
between the displacement fields indicated by the vertical 
label of each plot, as a function of the over-density smoothed on a 3 \hMpc scale.}
\label{fig:velAngles_Gauss}
\end{figure*}

Fig.~\ref{fig:velAngles_Gauss} is the same as 
Fig.~\ref{fig:velAmplitudes_Gauss}, but for 
the angle between the
smoothed components of the displacement fields. The measured angle is always
positive, but we made for clarity a reflection at the horizontal axis. 
To help understanding this figure one can assume for instance the  SIM IC
displacement direction to
be  chosen at random within a cone of angular size  10 (respectively 20) 
degrees centred
on the MAK/SIM NL displacement direction. In this case, the observed
distribution of angles would be bimodal and would peak at  6.7 
(respectively 13.3) degrees.
 
In regions of low to average densities, the quality 
of the MAK reconstruction of
the directions of the SIM NL displacement is striking. Above the mean density,
the scatter starts to increase as expected again because 
these zones roughly correspond to 
virialized regions at $z=0$, where reconstruction
is more difficult, but the bulk of the PDF remains within a $\pm$ 2 degrees 
offset. Note also that in the most strongly over-dense regions
the displacement amplitudes are typically smaller than in regions of lower
density, so that the consequences 
of significant errors in direction are not dramatic, simply because the 
magnitude of the displacement is small. On the other hand, comparing the
MAK/SIM NL to SIM IC displacements, we find that the agreement
is much less good, the most likely
difference between the two directions being larger
than $\sim 2.5$ degrees at the smoothing scales considered here. 

\subsection{Comparison of the ``density fields''}
\label{sec:ResultsGauss:Dens}

We estimate the {\it density contrast}, $\delta$, using the linear-regime relation
$\delta({\bf x}) \propto -\nabla_q\cdot {\bf v}$. 
For the SIM IC sample this is a fair estimate of the density field at this redshift.
For the SIM NL and MAK samples, this basically gives an estimate of the primordial  
density by simple application of the Zel'dovich approximation.

\begin{figure*}
\includegraphics[width=16cm]{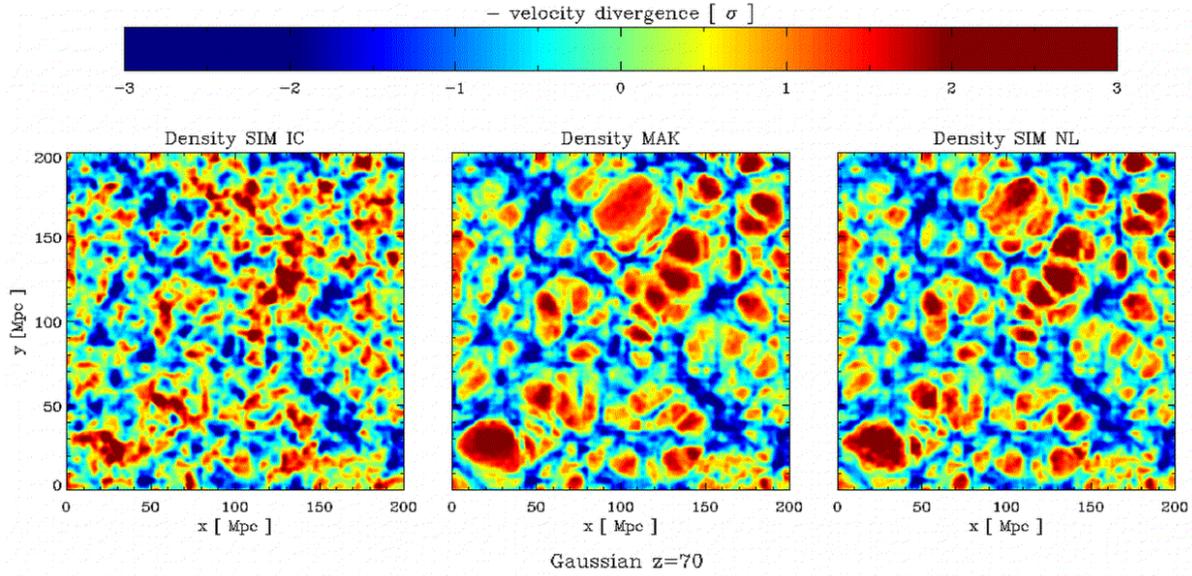}
\caption{Same as Fig.~\ref{fig:velz_z70_Gauss} but for the projected negative 
divergence of the displacement field (proportional to the over-density in the linear regime). 
MAK reproduces very well the SIM NL field, at all densities.
On the contrary, there are significant differences in regions of 
medium to high densities between the MAK
and the SIM IC fields: $\nu=1.5$ to 2 regions are more patchy 
in MAK and SIM NL than in SIM IC, and the sharp,
highest density peaks in SIM IC are not reproduced in the other two maps.}
\label{fig:dens_z70_Gauss}
\end{figure*}

\begin{figure*}
\begin{tabular}{@{}c@{}}
\includegraphics[width=8cm]{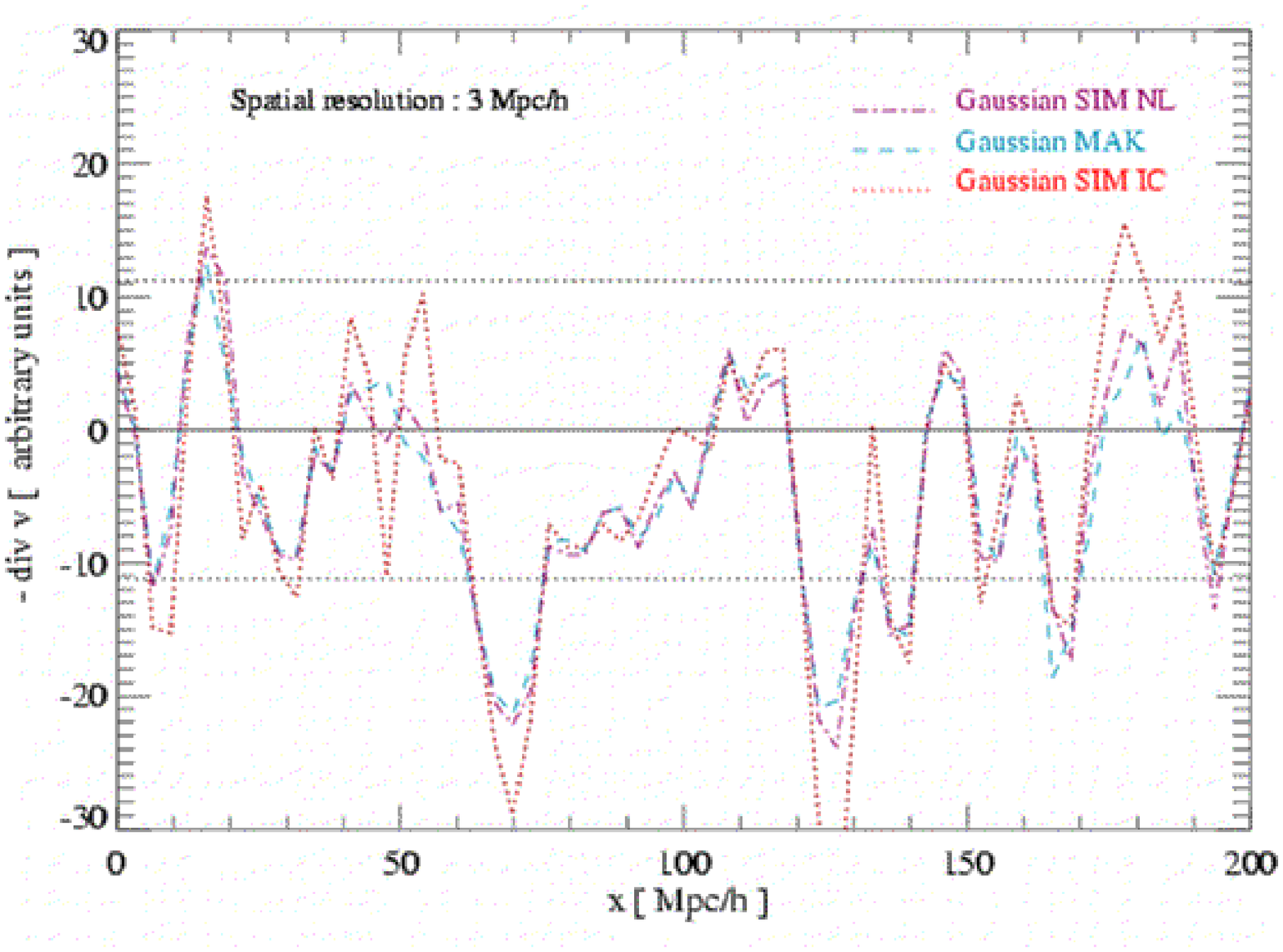} \\
\includegraphics[width=8cm]{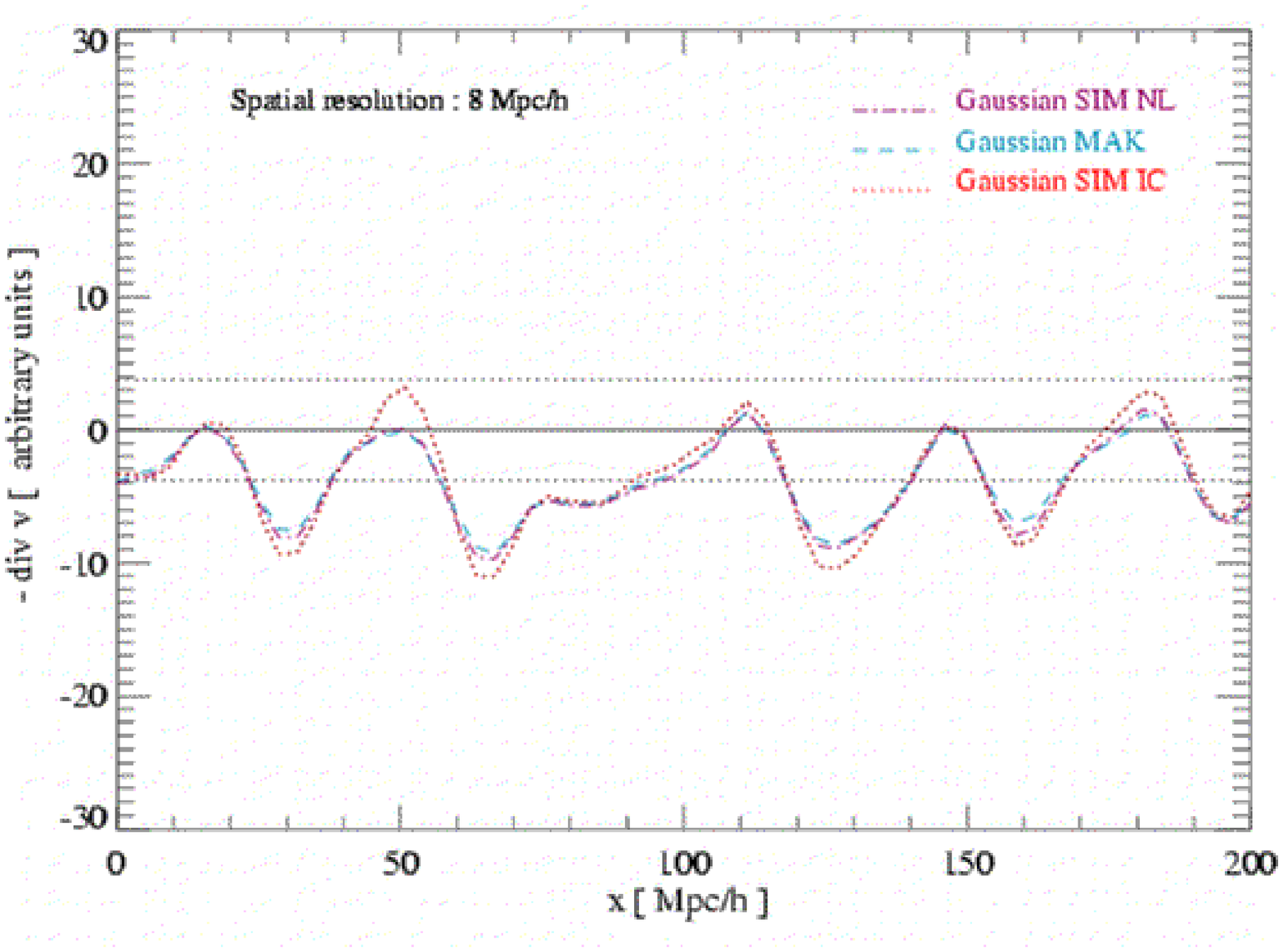}
\end{tabular}
\caption{
Same as Fig.~\ref{fig:TestLine_vz_Smooth0_Gauss}, but for the
divergence of the displacement 
fields. The resolution is 3 (resp. 8) \hMpc in 
the upper (resp. lower) panel. Again, MAK reproduces very
well the SIM NL displacement field, even though the curves are more 
noisy than in Fig.~\ref{fig:TestLine_vz_Smooth0_Gauss}, the consequence of taking
the derivatives.}
\label{fig:TestLine_divvel_Smooth0_Gauss}
\end{figure*}

In practice, to estimate $\nabla_q\cdot{\bf v}$, we first assign the
displacement field of the particle set $S_{1}$ to a $64^3$ grid using nearest grid
point interpolation (hereafter NGP) and then Fourier transform to take the gradient.
As a result of taking the derivative, the $\nabla_q\cdot\bf v$ field is intrinsically more noisy than the
displacement field. Fig.~\ref{fig:dens_z70_Gauss} is the pendant of Fig.~\ref{fig:velz_z70_Gauss}
but for $\nabla_q\cdot\bf v$. It is supplemented with Fig.~\ref{fig:TestLine_divvel_Smooth0_Gauss}, 
which shows the 3- and 8 \hMpc smoothed divergence field computed 
along a line parallel to the $x$ direction cutting through the centre of the box.

The MAK density field 
clearly reproduces the SIM NL field very well at all values of $\nabla_q\cdot\bf v$. 
On the contrary, the
MAK field only reproduces well the SIM IC field in regions of low to 
average densities. High-density ($\nu\sim1.5-2$)
patches are much more extended in MAK and in SIM NL than in SIM IC, and 
the $\nu\sim2.5$ to 3 levels of the SIM IC field are
barely reached by MAK or SIM NL. 

\begin{figure*}
\includegraphics[width=16cm]{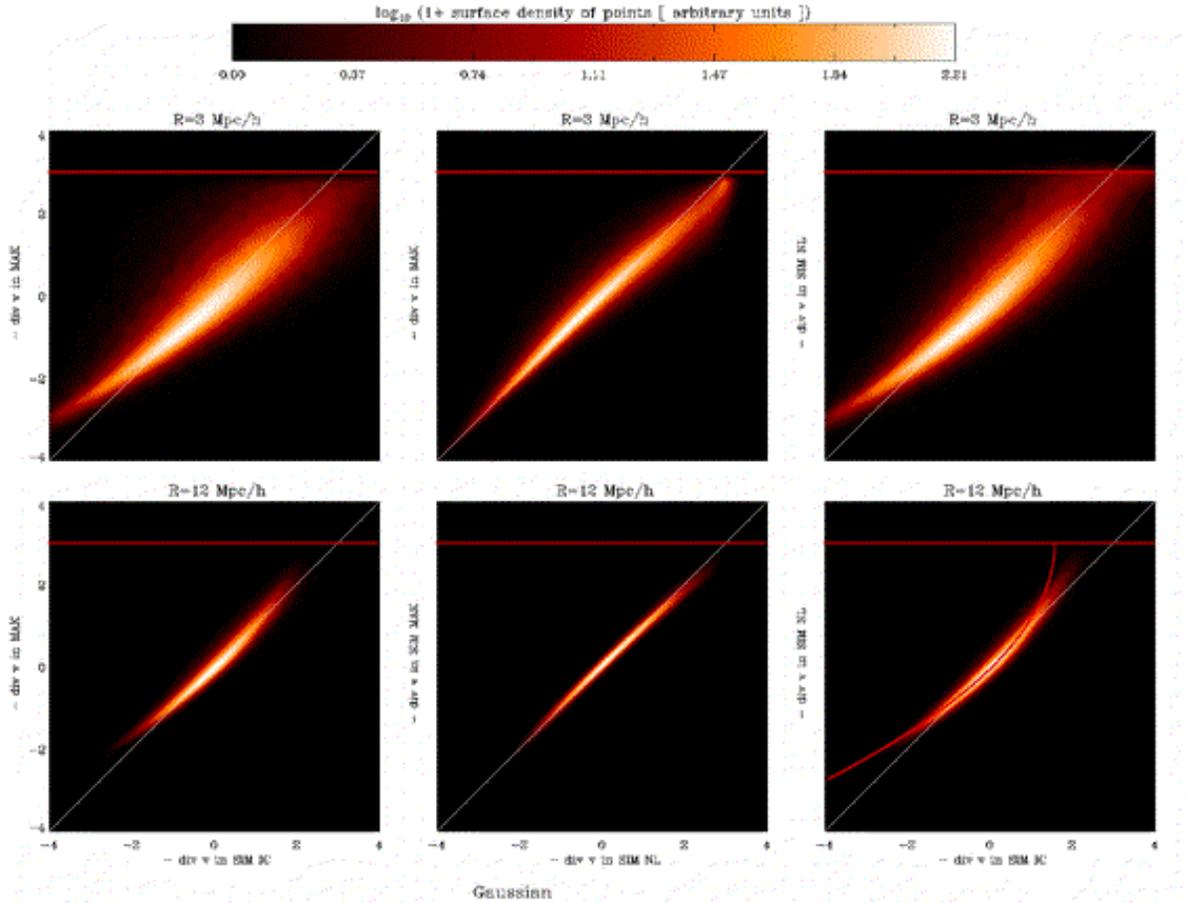}
\caption{Scatter plots comparing in a point-by-point 
perspective (nodes of a regular grid) the density of the MAK field to the
density of the SIM IC field (left panels), the MAK vs. SIM NL density fields (middle) 
and the SIM NL vs. SIM IC (right). The upper (resp. lower) panels are for
a 3 (resp. 12) \hMpc smoothing length. We have used the negative divergence of 
the displacement as a proxy for the density. We recover the expected  
tight correlation between MAK and SIM NL and the absence of any 
significant bias. A horizontal line is displayed on each panel. It corresponds to the expected
upper bound $-\nabla_q\cdot{\bf v}=3$ as discussed in the text. In the bottom right
panel, the curve shows the result given by the spherical collapse approximation [see
Eq.~(\ref{eq:simple})]. The limitations of this model, discussed later
in Section~\ref{sec:Prospects}, justify the 
choice of $12 \hMpc$ instead of $8\hMpc$ for the smoothing length.}
\label{fig:ScatterSmooth_Gauss}
\end{figure*}

This can be interpreted as follows. The fact that the simulated and the reconstructed 
displacement fields agree
very well in the under-dense regions in all the cases is not surprising:
indeed, the Zel'dovich mapping, hence SIM IC, is known, at 
least to some approximations, to work well in these regions
\citep{SS96} and MAK reconstructs perfectly the Zel'dovich
dynamics prior to the shell crossing.
Therefore, in the under-dense regions, agreement 
among SIM IC, SIM NL and MAK is expected at least qualitatively.
In detail, however, extra complications 
arise due to nonlinear dynamics which will be discussed below in more detail 
(see also Fig.~\ref{fig:ScatterSmooth_Gauss}). 

\begin{figure*}
\includegraphics[width=16cm]{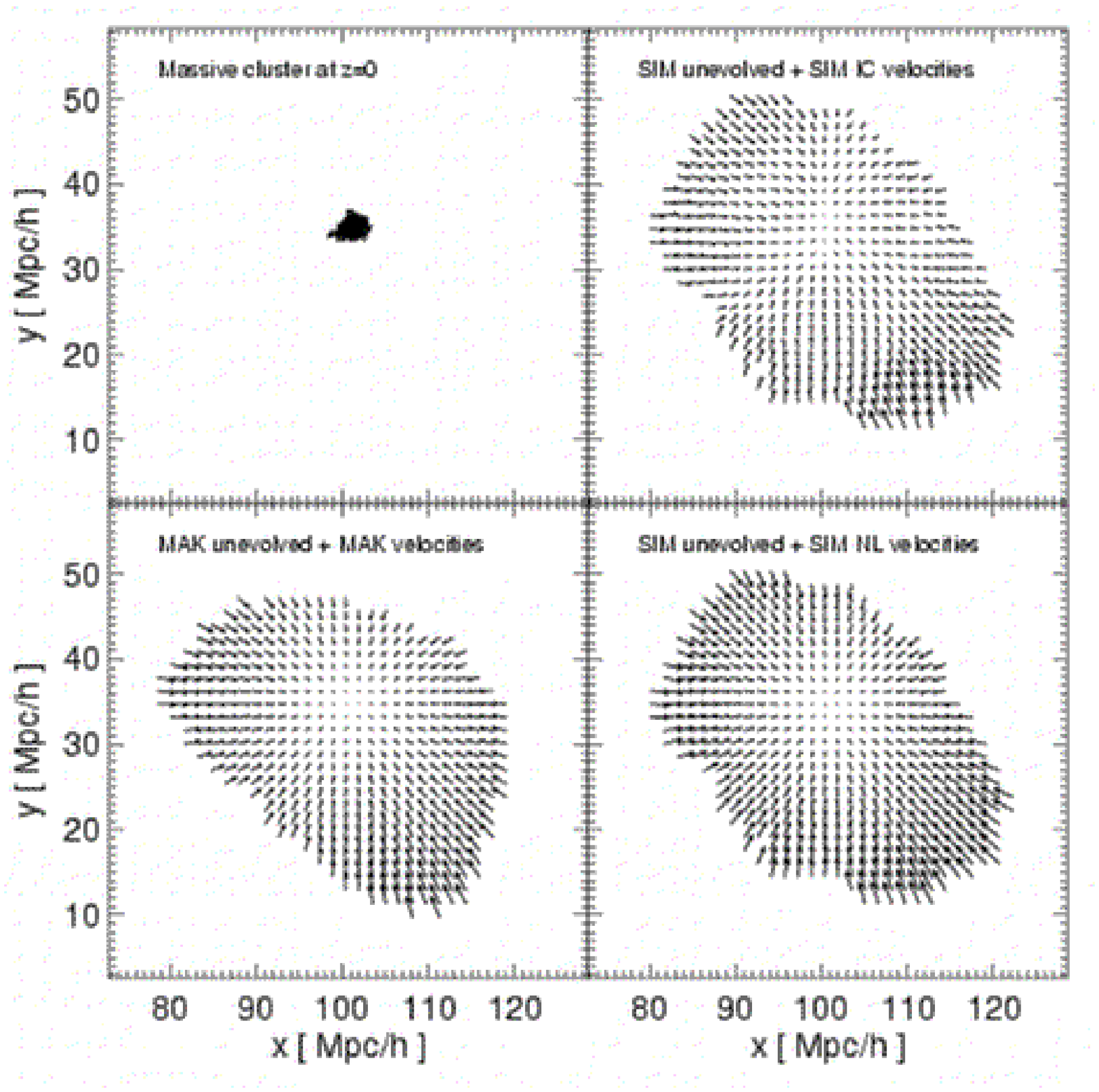}
\caption{
Upper left panel: z-projected simulated spatial distribution 
of particles of the third most massive cluster (with total mass
$M_{\rmn{tot}}=1.4\times10^{15} \hmsun$ and 4316 particles) of the Gaussian 
simulation at the present time. In this Figure we use the full resolution
analysis, {\it i.e.,} sample $S_0$. Upper right panel: 
z-projected Lagrangian region in the uniform grid for the particles ending
in the cluster and the Zel'dovich SIM IC velocities. 
Lower left panel shows the 
MAK-reconstructed Lagrangian region and MAK-reconstructed 
displacement field. Lower right panel shows the true Lagrangian region (as
in the top right panel) but with the SIM NL displacement field.
We use arbitrary units but
a common scale for all plots, so that the amplitudes can be compared. 
In the right top and right bottom panels we use the exact 
initial positions provided by
the simulation and in the left bottom panel 
the Lagrangian region reconstructed by MAK.
}
\label{fig:ClusterVelocField}
\end{figure*}

In the over-dense regions, the interpretation of the results is more complex:
the large patches with roughly constant
density in SIM NL and MAK correspond to collapsed objects at $z=0$, as 
explicitly illustrated by Fig.~\ref{fig:ClusterVelocField} where we 
analyse a typical cluster at full resolution, {\it i.e.} extracted from the $S_0$
sample. Using Fig.~\ref{fig:ClusterVelocField} as a
guideline, let us
consider a Lagrangian region that will collapse to an object of 
negligible size, as illustrated by the upper left panel. 
It is trivial to realise that the displacement from the initial 
to the final stages is purely radial as shown in the left and right
bottom panels for MAK and SIM NL respectively, and therefore that minus its divergence 
is equal to $3$, which is the upper bound expected for MAK reconstruction and
for the simulation (more approximately in the latter case).

We thus see that the left and right bottom panels of Fig.~\ref{fig:ClusterVelocField}
are extremely similar. They mostly differ in the boundary of the Lagrangian
region occupied by the cluster: this means that the main difference between
MAK and SIM NL relies on particles which have been assigned to the wrong
collapsed objects. Particles that are not in the intersection of
the two patches of the left and right bottom panels are clearly
assigned wrongly by MAK and will give totally wrong displacement fields.
We, however, notice that the boundaries of these "peak patches" are nearly the
same, which explains why MAK performs so well. At this point, it is important
to notice that exact recovery of initial positions of particles belonging to
the same collapsed object is not crucial and the
quality of the reconstruction of the displacement field is only marginally
affected if the positions of the particles ending up in the cluster are swapped 
by the MAK assignment. This demonstrates that there is no need to worry about
the occasional low level of "ideal" reconstruction presented in Table 1. 
Of course, the linear displacement field is not radial as shown in the upper right
panel of Fig.~\ref{fig:ClusterVelocField}, because it accounts for the
details in the distribution of
the initial fluctuations that are in the Lagrangian volume: in a hierarchical
model such as CDM, smaller structures form first which then merge and create
larger structures. In our picture, this would mean successive episodes of
approximately locally radial displacements. 

Clearly, Fig.~\ref{fig:dens_z70_Gauss} already emphasises the 
strengths and weaknesses of MAK:
it seems to reconstruct the nonlinear displacement field and its
divergence very well. However, the real goal is the recovery of the primordial
density field: in this case,
MAK performs qualitatively well in the under-dense regions but fails to do as well in 
the high-density regions due to the presence of collapsed structures, as
explained above.

\begin{figure*}
\includegraphics[width=8cm]{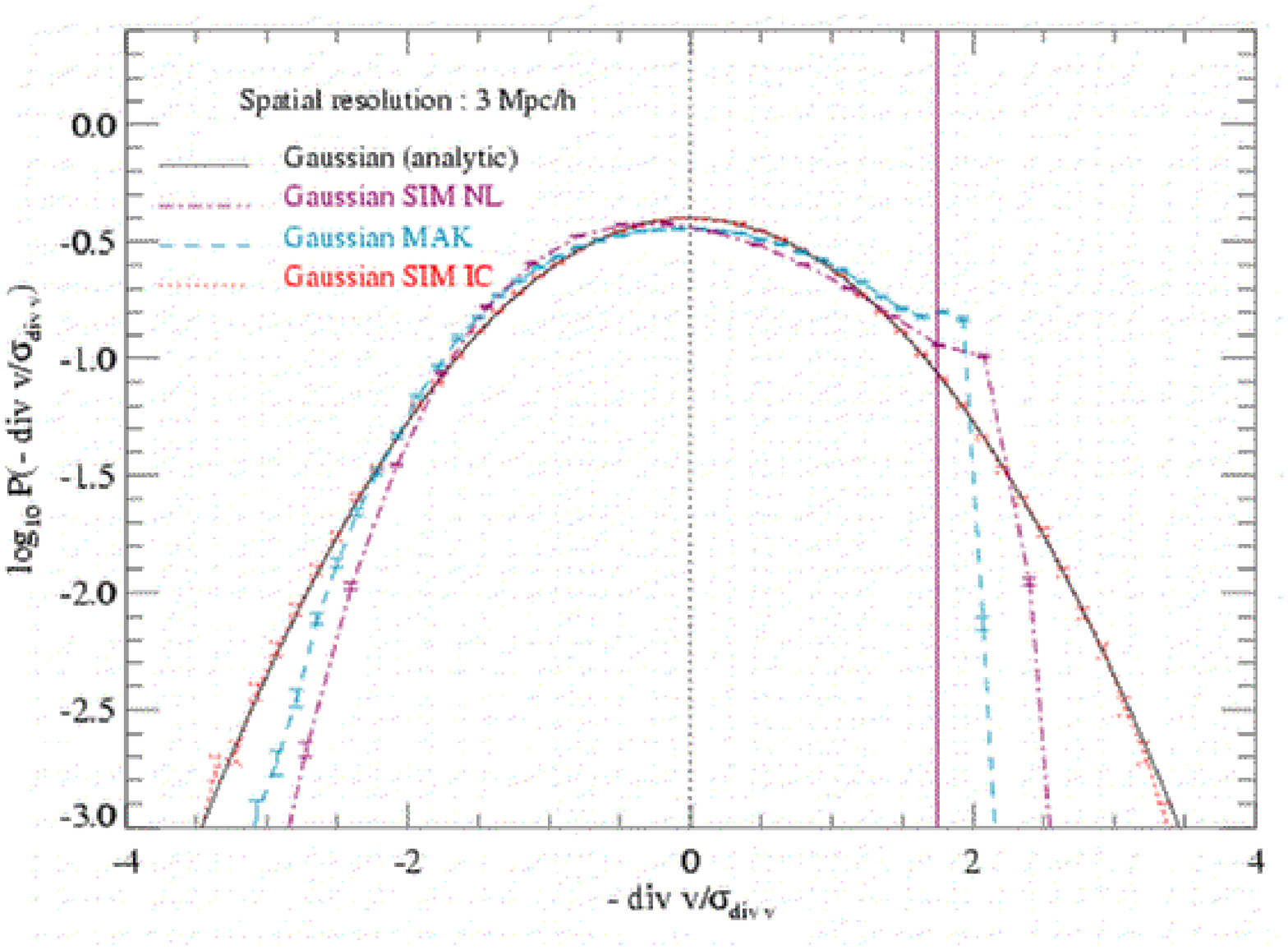}
\includegraphics[width=8cm]{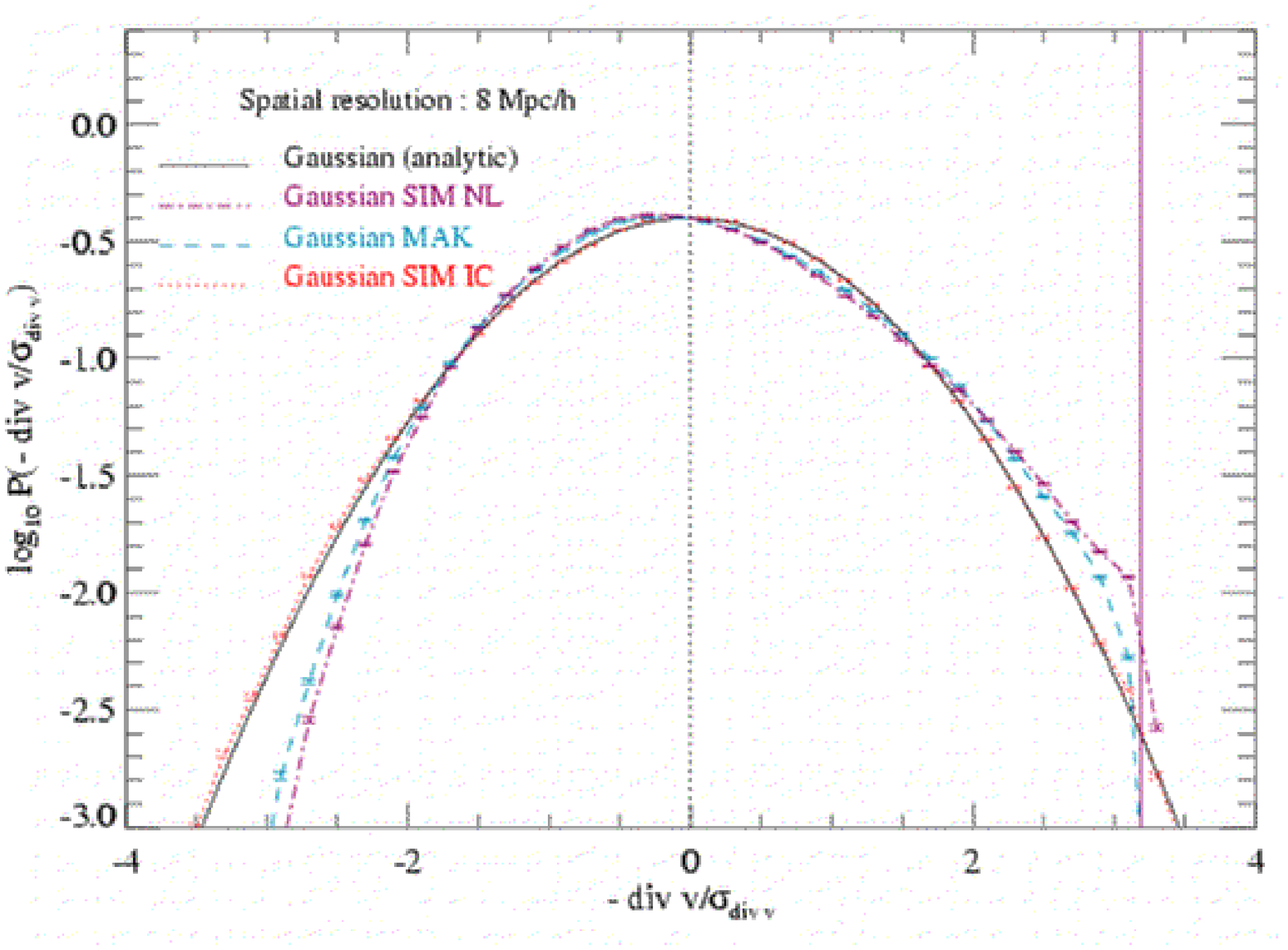}
\caption
{
Same as Fig.~\ref{fig:velz_PDF_Gauss} but for the divergence of the
displacement field. The left and right figures correspond 
to $3$ \hMpc and $8$ \hMpc smoothing scales respectively.
Both MAK and SIM NL which agree very well have a cutoff at large
value of $\nu$ indicated by the vertical line (see text for further explanations).
}
\label{fig:dens_PDF_Gauss}
\label{fig:dens_PDF_Gauss_3}
\end{figure*}

To illustrate these results in a more quantitative way,
Fig.~\ref{fig:dens_PDF_Gauss} examines the PDF of the divergence of the displacement. 
The upper and lower panels of the figure correspond to the field smoothed
here within a radius of 
$\sim 3$ \hMpc and $\sim 8$ \hMpc respectively.  Again, the PDF 
of the MAK density field  reproduces very
 well that of the SIM NL density field, including its positive skewness 
and its abrupt cutoff at large $\nu$ which corresponds to the present value of 
$-\nabla_q\cdot\bf v=3$ as discussed above.

Fig.~\ref{fig:ScatterSmooth_Gauss} goes into deeper 
detail by examining the bivariate distributions: it
represents
scatter plots of the divergence of the displacement field for MAK versus SIM IC
(left panels), MAK versus SIM NL (middle panels) and SIM NL versus SIM IC (right
panels) computed for the fields smoothed  at 3 and 12 \hMpc (upper and lower rows).
The banana-shape observed in the right panels is due to nonlinear clustering and
can be predicted from perturbation theory, similarly to what has previously 
been done for the divergence of the 
velocity field \citep{Bern94b,Bern99}. A calculation 
of $-\nabla_q\cdot {\bf v}$ relying on the spherical collapse 
model is given in Section ~\ref{sec:Prospects} and displayed on the bottom right panel.
If one neglects the cutoff expected for SIM NL at $-\nabla_q\cdot \bf v\sim 3$
the nonlinear gravitational effects increase the value of $-\nabla_q\cdot \bf v$
in the under-dense regions and increase it in the over-dense regions compared
to the linear predictions.
We notice here that the qualitative results stated previously about under-dense
regions need to be quantified: even in the low density regime the recovery of the
initial density field from the nonlinear displacement field can be nontrivial
without further priors.
In the middle panels, we see that MAK again agrees extremely well with SIM NL
with a very small bias:
the scatter around the diagonal is much smaller than in the right
panels although we see, in the $3$ \hMpc smoothing case, an increase of
the dispersion for positive values of $-\nabla_q\cdot {\bf v}$ and a 
slight bias at larger values
of $-\nabla_q\cdot{\bf v}$.  This
comparison between SIM NL and MAK confirms fully the nontrivial nonlinear
nature of MAK reconstruction. The left panels are very similar to 
the right panels, as expected.

\subsection{Present-day peculiar velocity fields}
\label{sec:ResultsGauss:PecVels}

An important and direct outcome of MAK reconstruction, the peculiar velocity
field ${\bf v}_{\rmn{MAK}}({\bf x})$, is a surprisingly good approximation  to the
present day true peculiar velocity field ${\bf v}_{\rmn{pec}}({\bf x})$.

\begin{figure*}
\includegraphics[width=16cm]{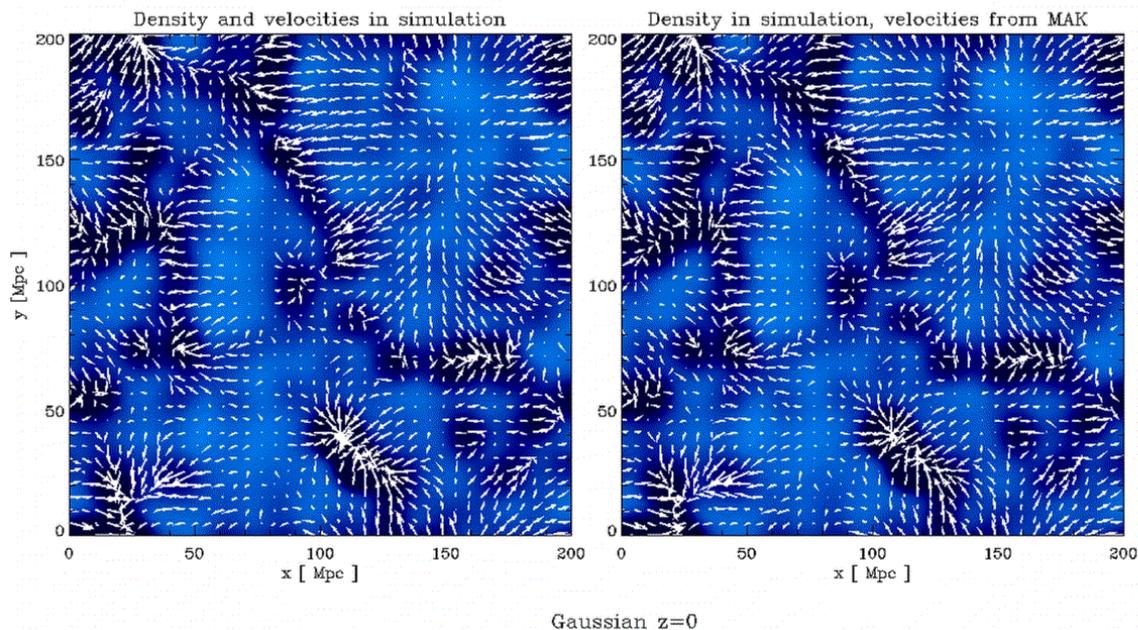}
\caption{$z=0$  $v_{\rmn{x}}$-$v_{\rmn{y}}$ velocity field of 
a 20 \hMpc-thick slice normal to the z-direction
cutting at a depth of $\sim 1/3$ of the simulation 
box in z through the Gaussian simulation; the 
left and right panels respectively correspond to
the simulated (peculiar) velocity field and to the reconstructed 
velocity field. The  8 \hMpc smoothing scale guarantees
that we are not affected by shell crossing.  Underlying 
the velocity field we show in colour map 
the 3-dimensional density field at $z=0$ cut from
the same z-slice and projected in z (it is the same density field 
in the left and right panels, and black is over-dense). There is 
quasi-perfect agreement between the two velocity fields.}
\label{fig:vel_z0_Gauss}
\end{figure*}

${\bf v}_{\rmn{MAK}}({\bf x})$ for a given particle is obtained simply by
scaling of the MAK displacement obtained previously at
reconstructed particle's position (node of the Lagrangian grid), and assigning 
the displacement to this particle's position at $z=0$
following equations~(\ref{eq:Zeldovich}) and~(\ref{eq:PosScaling}). It is unclear, except
maybe at very large scales, that ${\bf v}_{\rmn{MAK}}({\bf x})$ can approximate 
${\bf v}_{\rmn{pec}}({\bf x})$. Fig.~\ref{fig:vel_z0_Gauss}
shows that this is indeed the case on scales of $8$ \hMpcDot There, we show
the $z=0$ $v_{\rmn{x}}$-$v_{\rmn{y}}$ velocity field extracted at $z=0$ from
a 20 \hMpc-thick slice normal to the z-direction and cutting 
through the centre of the Gaussian simulation. (We use the $S_{1}$ sample 
in this Figure.) The left (resp. right) panel shows the ${\bf v}_{\rmn{pec}}({\bf x})$ field
given by the simulation (resp. ${\bf v}_{\rmn{MAK}}({\bf x})$), smoothed with a top-hat kernel of radius
8 \hMpcDot The underlying colour map gives the simulated dark matter density
field smoothed on the same scale and then projected; it is the same for 
both panels. On these scales, we find excellent agreement between the 
simulated peculiar velocities and the reconstructed velocities.

\begin{figure*}
\includegraphics[width=6.9cm]{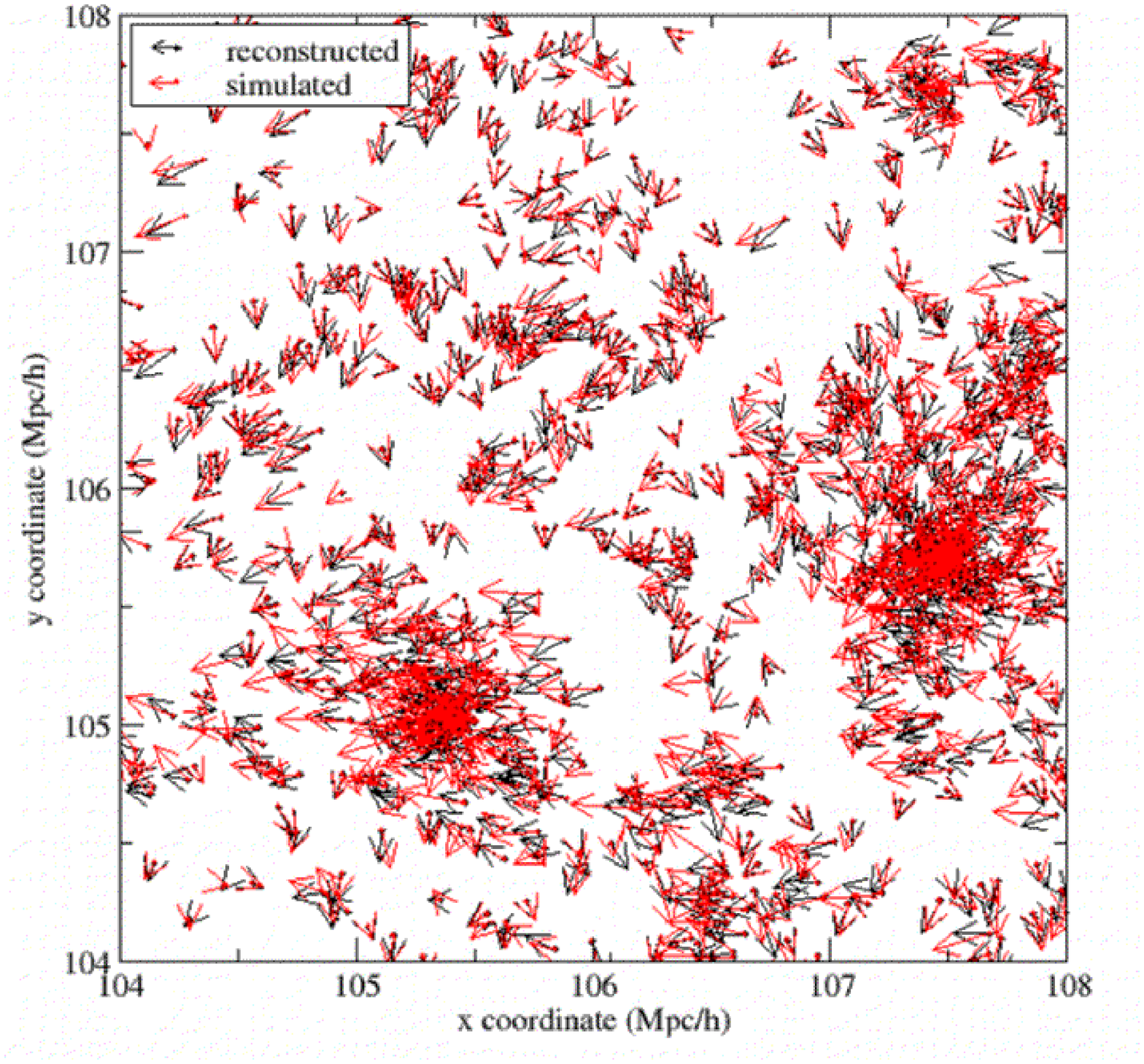}\qquad\quad
\includegraphics[width=8.5cm]{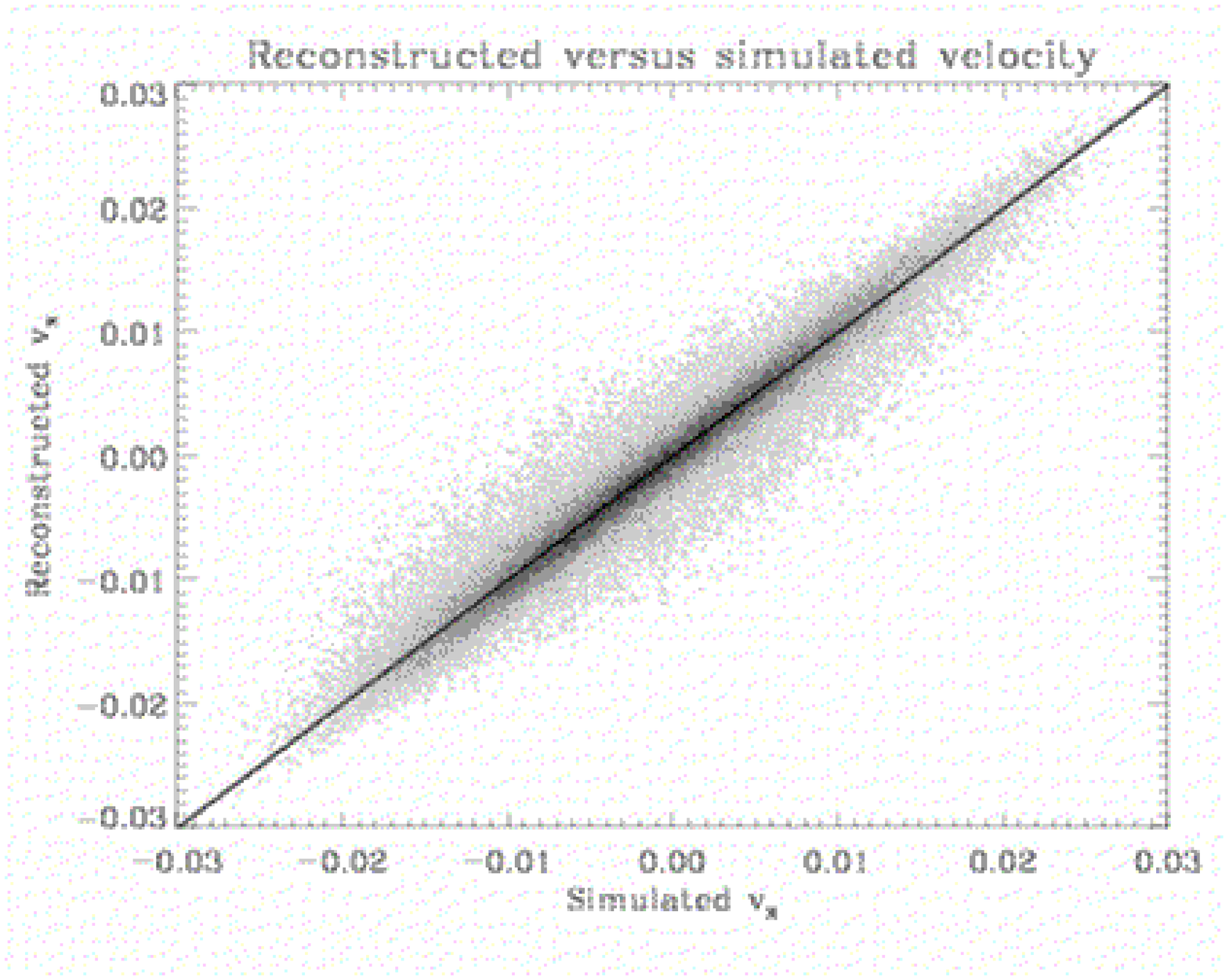}
\caption{{\it left panel: }$z=0$  $v_{\rmn{x}}$-$v_{\rmn{y}}$ velocities of particles 
selected from a 4 \hMpc-side and 20\hMpc deep sub-volume taken from the Gaussian 
simulation. Simulated peculiar (resp. reconstructed)
velocities are shown with the red (resp. black) arrows. There is 
fairly good agreement in the voids, but more noticeable differences
 in the outskirts of haloes.
{\it right panel: } scatter plot of the
reconstructed and simulation velocities at z=0 for the Gaussian
sample of $64^3$, sample Gaussian $S_1$ of Table 1, demonstrates closely the
goodness of the reconstructed velocity field. Axes are in box units (one unit
is 200 \hMpc). 
}

\label{fig:vel_z0_Gauss_SmallScale}
\end{figure*}

On much smaller scales, Fig.~\ref{fig:vel_z0_Gauss_SmallScale} details 
the simulated and reconstructed present-day
($v_{\rmn{x}}$,$v_{\rmn{y}}$) components of ${\bf v}_{\rmn{MAK}}({\bf x})$ 
and  ${\bf v}_{\rmn{pec}}({\bf x})$ for particles
 selected in a $4^3$ \hMpc region of the Gaussian 
simulation. (We use the $S_{0}$ sample here.) Although the agreement in the
sparse/under-dense regions is still generally very good, differences arise 
between the simulated and the reconstructed particle 
velocities in regions of medium and higher densities.

A thorough comparison between MAK and
other methods of proper-velocity evaluation 
({\it e.g.} Yahil et al 1991 ; Valentine et al. 2000 and Shaya
et al. 1995) remains to be done in future works where we plan to apply MAK to
mock and large-scale redshift galaxy catalogues.

The very good agreement on scales of $\sim$ 8 \hMpc suggests a
possible use of MAK to turn present-day redshift catalogues into
real space catalogues, in an iterative process to deal 
with redshift space distortion.  In applying MAK to real or mock
galaxy catalogues, a quantitative error analysis is surely needed to access the 
goodness of MAK in evaluation of quantities such as cosmological parameters or
mass to light ratio (some of these issues have been discussed in Mohayaee and
Tully 2005 for the application of MAK to NearBy Galaxies (NBG) catalog).

\subsection{Possible resolution issues}
\label{sec:ResultsGauss:ResolutionCheck}

Our study of the resolution effects consisted of analysing the full resolution
sample ($S_0$) and dense sample ($S_2$) [see Section \ref{sec:Method:PartSample}
and Table 1] and of comparing the
results to the sparse sample ($S_1$) used for most of the analyses up to now. 
We did not notice any significant 
differences between the full and sparse resolutions. 

\begin{figure*}
\includegraphics[width=8cm]{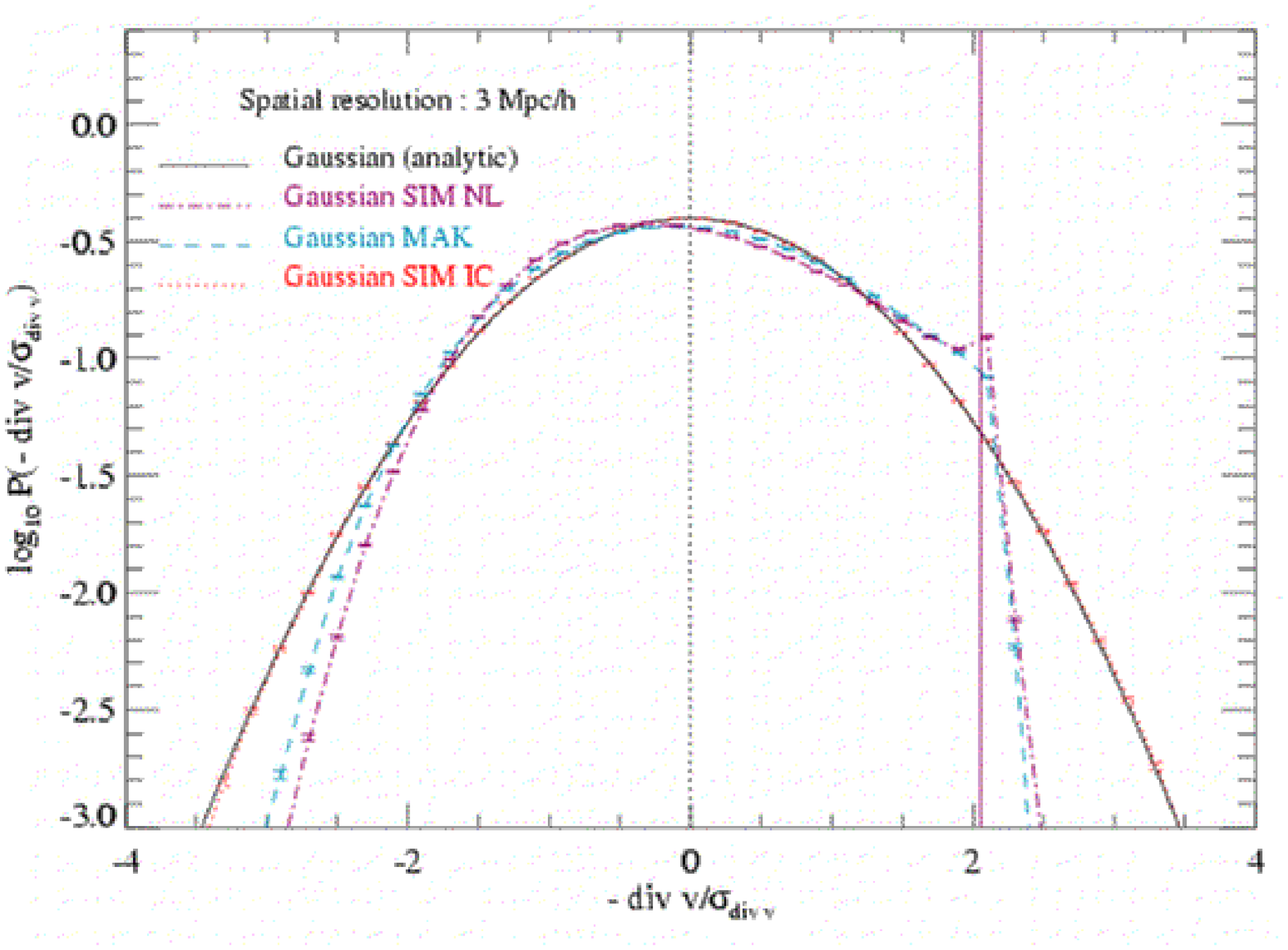}
\includegraphics[width=8cm]{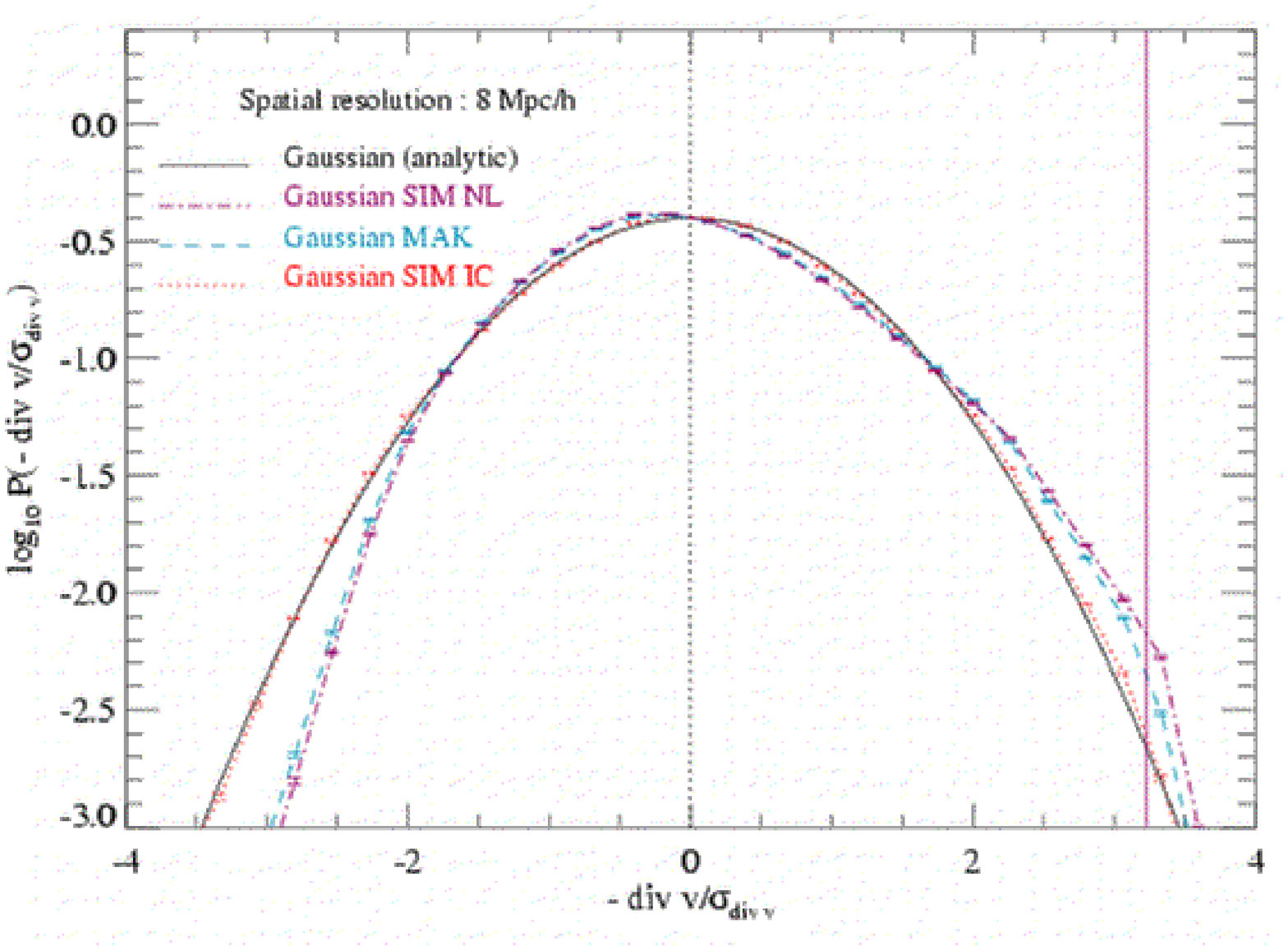}
\caption{
Same as Fig.~\ref{fig:dens_PDF_Gauss_3} but for the full set $S_{0}$
of $128^3$ particles rather than its sparse sample, $S_{1}$. 
}
\label{fig:dens_PDF_128Gauss}
\label{fig:dens_PDF_128Gauss_3}
\end{figure*}

To illustrate our point, the top and bottom panels of Fig.~\ref{fig:dens_PDF_128Gauss} 
compare the PDFs of SIM IC, MAK and SIM NL density
fields smoothed with a top-hat kernel of radius 3 and 8 \hMpc respectively using 
the full set $S_{0}$ of $128^3$ reconstructed particles.
There are no significant differences with respect to the PDF 
of Fig.~\ref{fig:dens_PDF_Gauss} computed for the set $S_{1}$.
Note that this nice agreement between the sparse and full samplings 
is further supported by the right panel of 
Fig.~\ref{fig:Scatter_128Gauss} and the last column of Table 1:
the "ideally" reconstructed Lagrangian positions in $S_1$ are consistent
with the fraction of the particles in $S_0$ that are reconstructed with accuracy
better than two mesh sizes.

The dense sample, $S_2$, tests as well, to some extent, the importance
of tidal and edge effects on the reconstruction. 
In practice, this can have important consequences for 
reconstructions of volume-limited (in the literal sense) catalogues which might 
not extend well above the scale of homogeneity.
For the dense sample
size considered here, $100$\hMpc, these effects seem to be negligible. 
For instance, the fraction of "ideally" reconstructed particles is $17\%$ to be
compared with the $18\%$ obtained for the full sample, $S_0$, reconstruction.

Note that we have performed these resolution tests as well for the $\chi^2$
model and the dense sample for the PVM and have obtained similar results.

\section{Reconstructing non-Gaussian density fields}
\label{sec:ResultsNonGauss}

In this section, we examine different non-Gaussian initial conditions and verify if
the results obtained previously for the Gaussian case still hold. We shall
see that this is indeed the case, namely that the displacement fields reconstructed
by MAK match extremely well their true nonlinear counterparts given by the simulations.

However, the real goal here is to study the properties of the initial density
field or in other words the divergence of the initial displacement field which
differs from the total final field, that is best reconstructed by MAK:
the nonlinear contribution due to gravitational collapse should be
subtracted from the reconstructed displacement field 
to allow tests for primordial non-Gaussianity.
This will be confirmed by the subsequent analyses which will demonstrate that
a ``simple" use of MAK reconstruction is insufficient for  
recovery of a small level of non-Gaussianity.
By ``simple" we mean the extrapolation of the nonlinear displacement field to early epochs using
the Zel'dovich approximation to infer the initial density field.

To carry out our analyses and reach these conclusions, we have explored the large
spectrum of non-Gaussian models, detailed in Section \ref{sec:Models}:
\begin{enumerate}
\item The $\chi^2$ model, examined in
  Sec. \ref{sec:ResultsNonGauss:ChiSquare}, which is strongly non-Gaussian but with the same
  topological properties as its Gaussian seed;
\item The primordial voids model (PVM), studied 
in Sec. \ref{sec:ResultsNonGauss:PVM}, which is strongly non-Gaussian
and initially inhomogeneous;
\item The weakly non-Gaussian Quadratic ($Q$) models, discussed 
in Sec. \ref{sec:ResultsNonGauss:Q}.
\end{enumerate}

\subsection{$\chi^2$ model}
\label{sec:ResultsNonGauss:ChiSquare}

\begin{figure*}
\includegraphics[width=16cm]{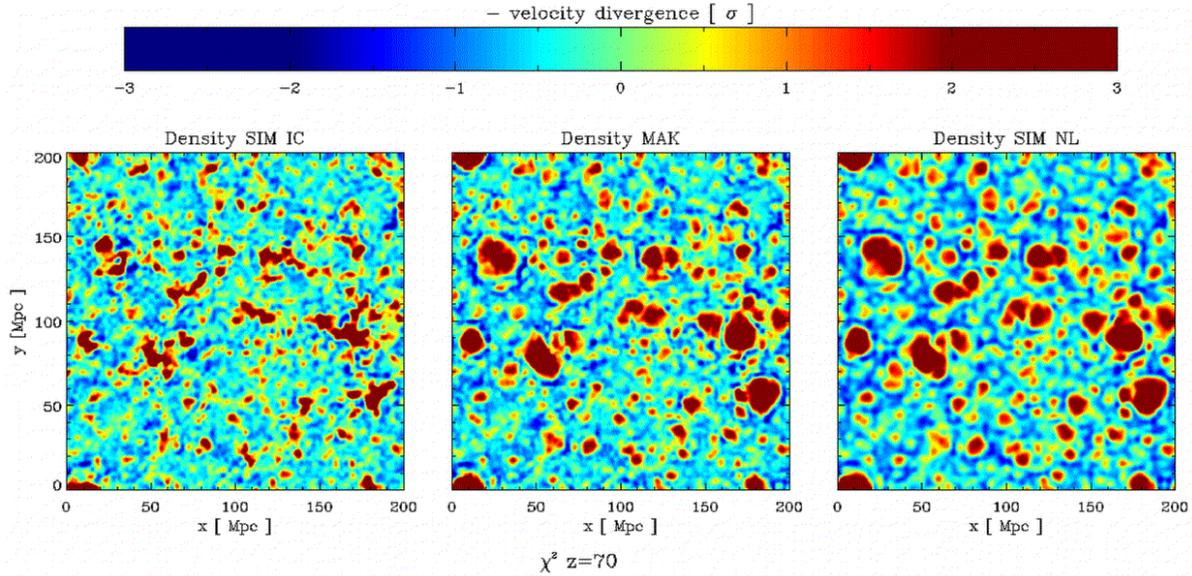}
\caption{Same as lower panel of Fig.~\ref{fig:dens_z70_Gauss}, but for the $\chi^2$ model.}
\label{fig:dens_z70_Chi2}
\end{figure*}

\begin{figure*}
\includegraphics[width=8cm]{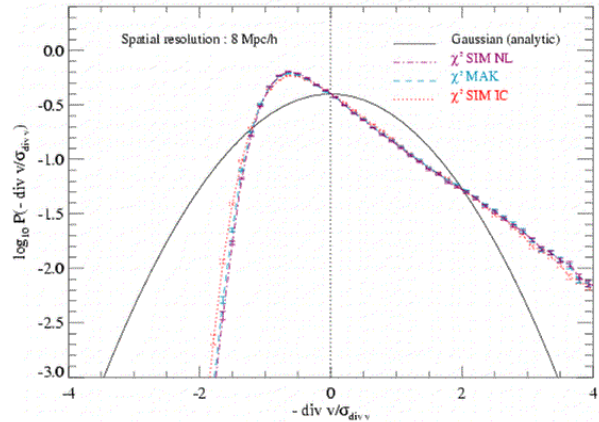}
\caption{Same as Fig.~\ref{fig:dens_PDF_Gauss} but for the $\chi^2$ model.}
\label{fig:dens_PDF_Chi2}
\end{figure*}

In Fig.~\ref{fig:dens_z70_Chi2} we examine $-\nabla\cdot\bf v$ in
a 10 \hMpc thick slice, similarly as in Fig.~\ref{fig:dens_z70_Gauss}, while
Fig.~\ref{fig:dens_PDF_Chi2} compares the PDF of $-\nabla\cdot{\bf v}$ for
SIM IC, MAK and SIM NL. The measurements are performed using a
top-hat smoothing length of 8 \hMpcDot The PDFs of
MAK and SIM NL agree perfectly, which is consistent with the 
visual inspection of Fig.~\ref{fig:dens_z70_Chi2}. They differ only slightly
from SIM IC PDF.
This might be surprising at first sight, because visually SIM NL and MAK look
rather different from SIM IC, at least 
for large values of $-\nabla\cdot{\bf v}$.  As explained in
length in our study of the Gaussian model, this difference is mainly due to the
presence of collapsed objects which induce an abrupt cut-off in the tail of the PDF for
large values of  $-\nabla\cdot{\bf v}$ (This cut-off lies outside the plot
range in Fig.~\ref{fig:dens_PDF_Chi2}).
However, a more careful inspection of
Fig.~\ref{fig:dens_z70_Chi2} shows a much better agreement in regions of lower
density, which explains the very good match between all of the three PDFs.

Except for the abrupt cut-off, the {\it small} difference as compared to the
Gaussian case between the MAK/SIM NL PDF and SIM IC is 
due to two factors. First, the contribution from initial non-Gaussianity is
strong enough to completely dominate over 
the contribution from gravity. Second, in this
model the effect of gravitational clustering is less intense than in the
Gaussian model because of the lower value of $\sigma_8$.
A noticeable consequence of this lower normalisation is that there are less
shell crossings and therefore the fraction of "ideally" reconstructed
particles is much higher than for the Gaussian model (see last column of Table 1).

\subsection{{\it PVM} model}
\label{sec:ResultsNonGauss:PVM}

\begin{figure*}
\includegraphics[width=16cm]{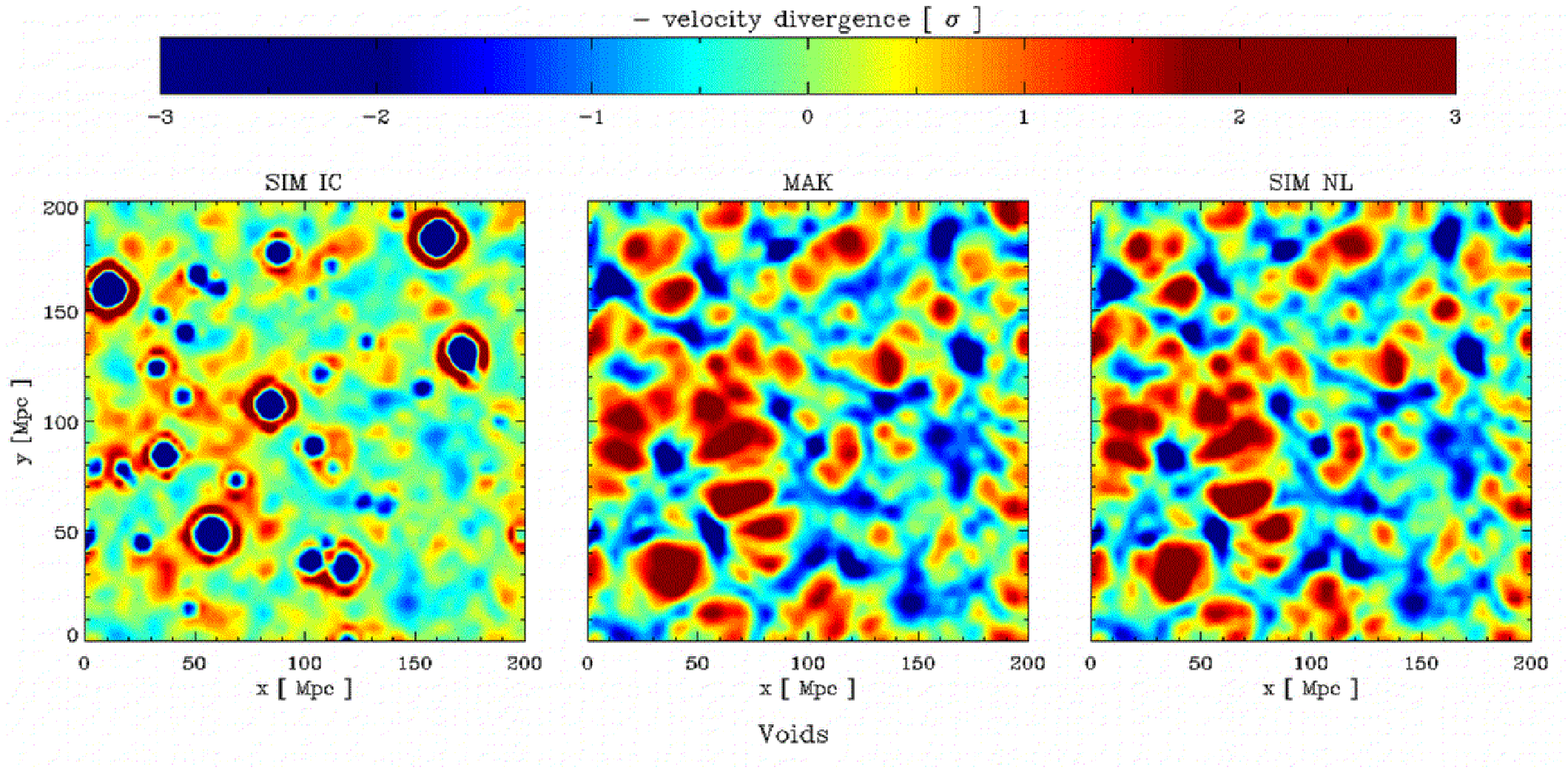}
\caption{Same as Fig.~\ref{fig:dens_z70_Chi2}, but for the {\it PVM} model. 
Again, the agreement between the MAK and SIM NL fields is very good.}
\label{fig:dens_z70_PVM}  
\end{figure*}

\begin{figure*}
\includegraphics[width=8cm]{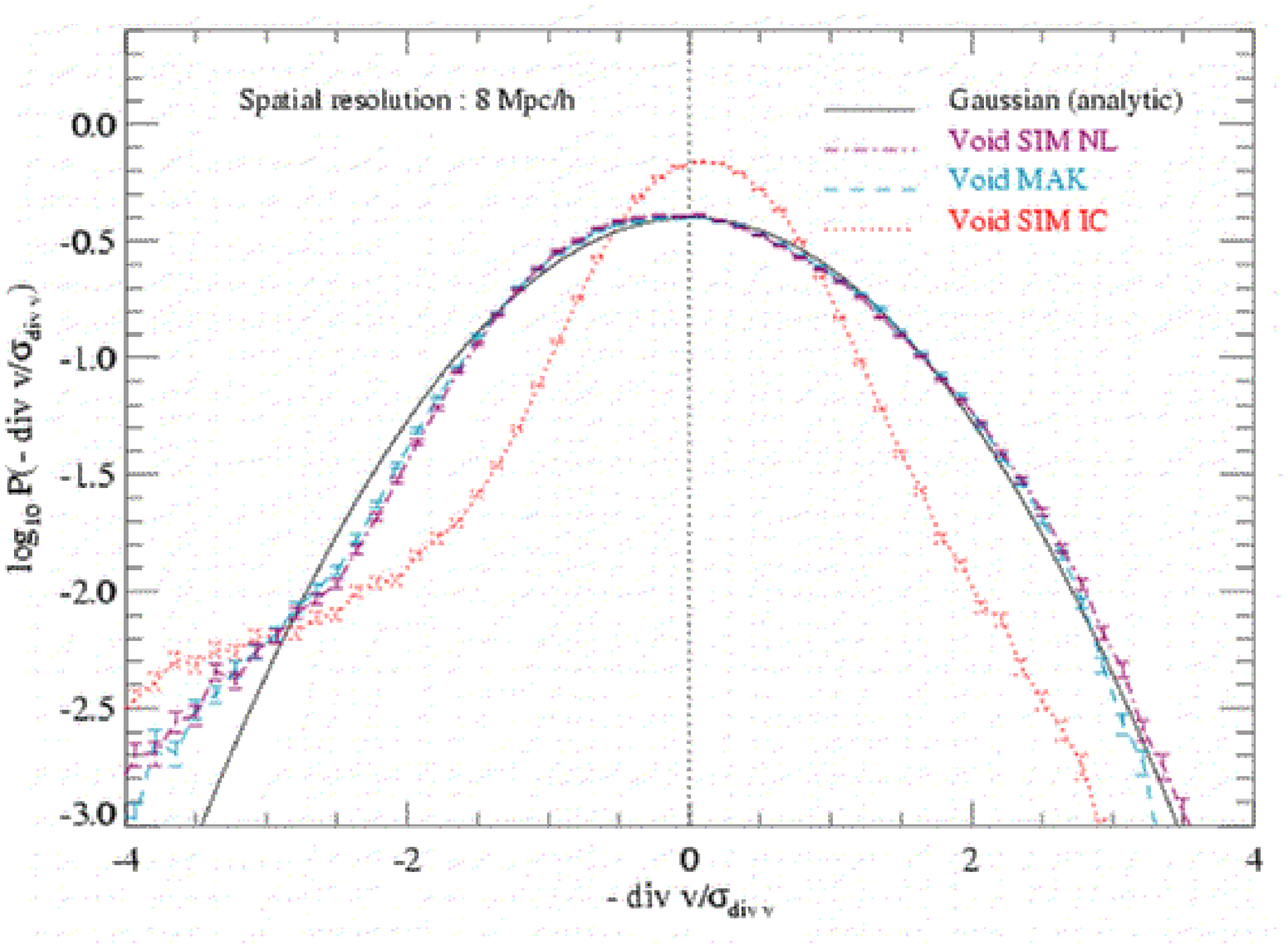}
\caption{Same as Fig.~\ref{fig:dens_PDF_Chi2} but for the {\it PVM} model. 
}
\label{fig:dens_PDF_PVM}
\end{figure*}

In Fig.~\ref{fig:dens_z70_PVM} we examine $-\nabla \cdot {\bf v}$,
 in a 10 \hMpc thick slice, similarly as in Fig.~\ref{fig:dens_z70_Gauss}, while
Fig.~\ref{fig:dens_PDF_PVM} compares the PDF of $-\nabla \cdot {\bf v}$ for
SIM IC, MAK and SIM NL. The measurements are performed with a
top-hat smoothing length of 8 \hMpc. 

Recall that for {\it PVM}, initial conditions are strongly inhomogeneous
at very high redshifts. As a result, it is in principle meaningless
to use MAK reconstruction. However, at some point, the initial conditions had
to be almost homogeneous, prior to the seeding of the primordial
voids. Hence, we can still assume that we are starting from Lagrangian
positions setup on a homogeneous grid. In our case, this homogeneous grid
is the one where particles are located prior to the Zel'dovich displacement and 
void creation as explained in Section \ref{sec:Models:PVM}. With this in mind,
a fair comparison between the
middle and right panel of Fig.~\ref{fig:dens_z70_PVM} can be made
and again the visual agreement between MAK and SIM NL is excellent. 
Expectedly, due to the early strong nonlinearities induced by voids creation and
subsequent shell crossings, the fraction of ``ideally" reconstructed
particles is rather low (28\%, see Table 1).

From the previous arguments, comparing MAK/SIM NL to SIM IC
has to be done with care. Clearly the strong ``density contrasts'' observed around
voids in the left panel are expected to be smeared out in the middle and right panels. 
Keeping this in mind, we notice that there is a global large scale qualitative
agreement between the three panels of Fig.~\ref{fig:dens_z70_PVM}, 
including the positions and extensions of the under-dense regions, although the
amplitudes of the fluctuations are different. 
This means that our nonlinear displacements fields have kept at least some
informations on the non-Gaussian nature of initial conditions even if the
middle and right panels differ significantly from the left one.

The results of the previous discussion are nicely illustrated by
Fig.~\ref{fig:dens_PDF_PVM} that we can now comment on in detail.
The bell-shaped part of the SIM IC PDF corresponds to the initial
Gaussian field generated prior to the seedings of the voids. 
The strong over-dense and under-dense regions 
appearing during the creation of the voids increase
the variance of the density field by adding tails to the PDF both
in the high and the low parts. This explains why the bell-shape
of the SIM IC PDF on
Fig.~\ref{fig:dens_PDF_PVM} 
is in fact narrower than a pure Gaussian with the same variance,
given the choice of representation, in units of 
$-\nabla_q\cdot {\bf v}/\sigma$. Note that this additional contribution
to the variance is, at least in the linear regime, a pure transient, as discussed
furthermore at the end of Section~\ref{sec:Prospects}.
If one neglects additional non-Gaussianity brought about by gravitational instability, one
thus expects that the bell-shaped part of the PDF converges to the pure Gaussian
at late times, as one can indeed observe in Fig.~\ref{fig:dens_PDF_PVM}.

The MAK and SIM NL PDFs  agree with each other very well as expected. In the 
positive $-\nabla_q\cdot {\bf v}$ part, nonlinear gravitational clustering and subsequent
shell crossings destroy information related to the strong over-dense spheres
around the voids. As a result, in this regime the shape of the PDF is 
qualitatively very similar to what would be obtained from the Gaussian case.
However, as noticed while examining Fig.~\ref{fig:dens_z70_PVM}, the
informations on the primordial voids is still present in 
the tail for the low values of $-\nabla_q\cdot {\bf v}$. Clearly, despite the fact
that SIM NL/MAK PDFs differ 
enormously from SIM IC PDF, they have still preserved
a detectable signature of primordial non-Gaussianity.

\subsection{Quadratic models}
\label{sec:ResultsNonGauss:Q}

\begin{figure*}
\begin{tabular}{@{}cc@{}}
\includegraphics[width=8cm]{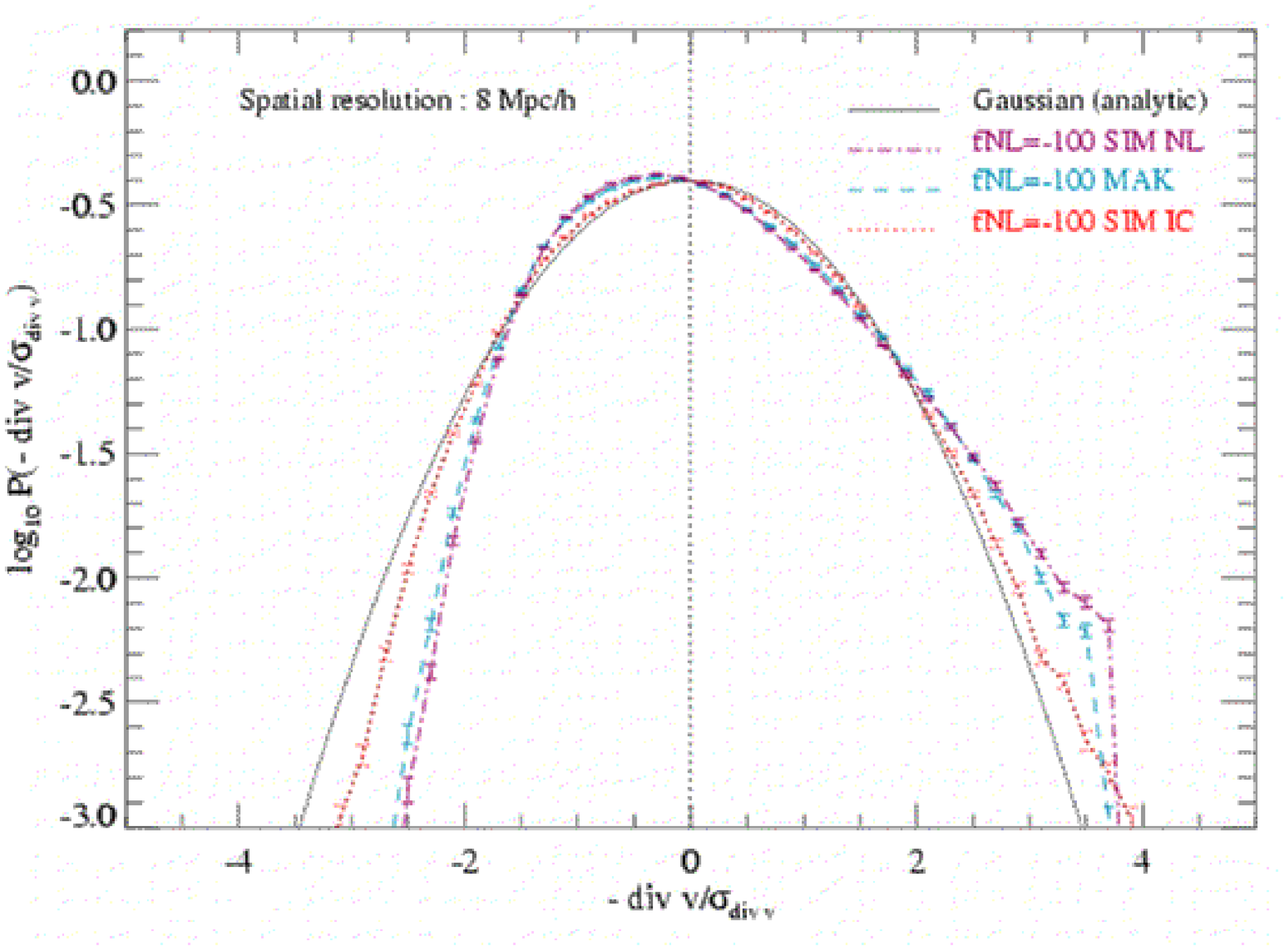}
& 
\includegraphics[width=8cm]{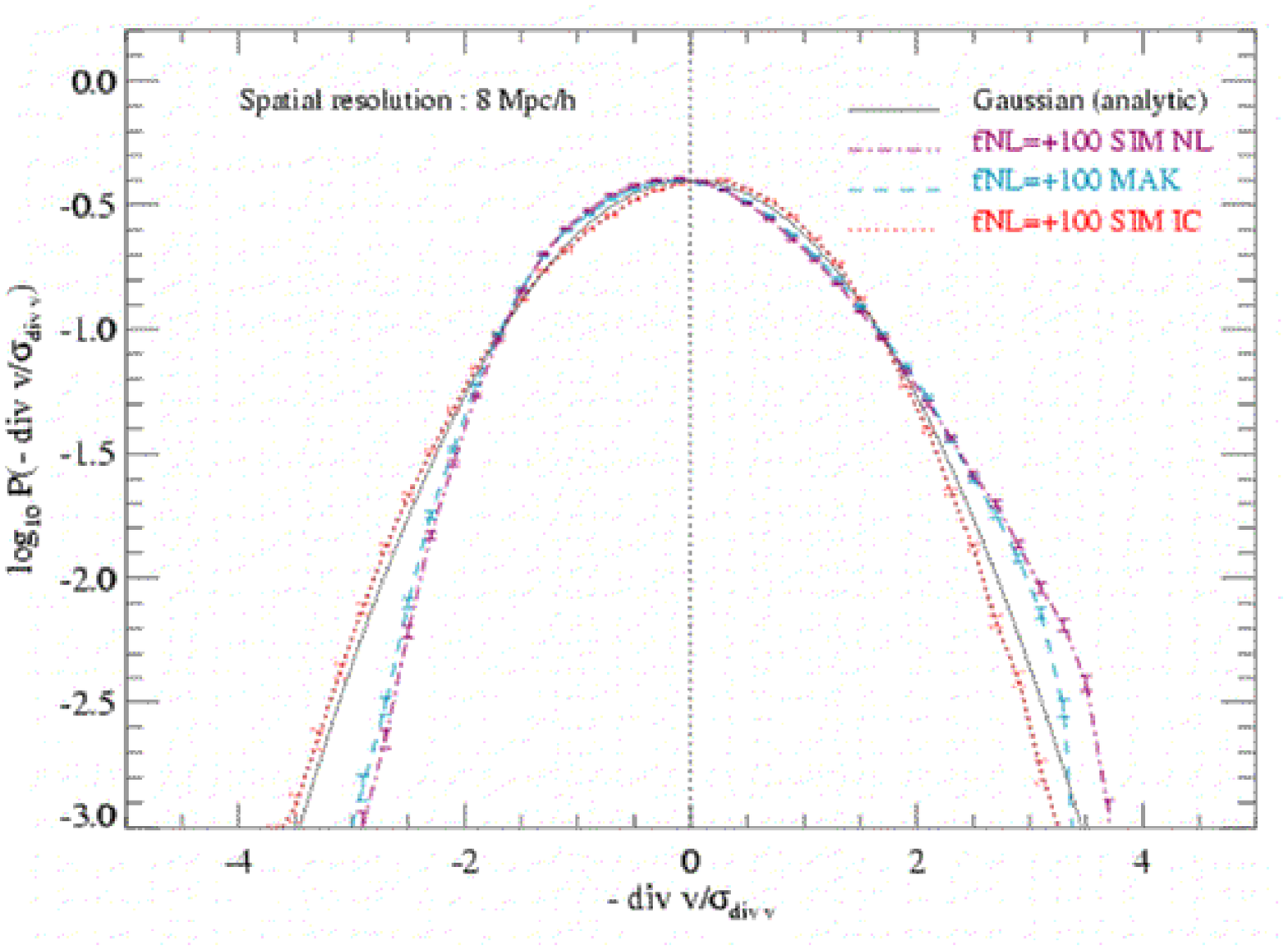} \\
\includegraphics[width=8cm]{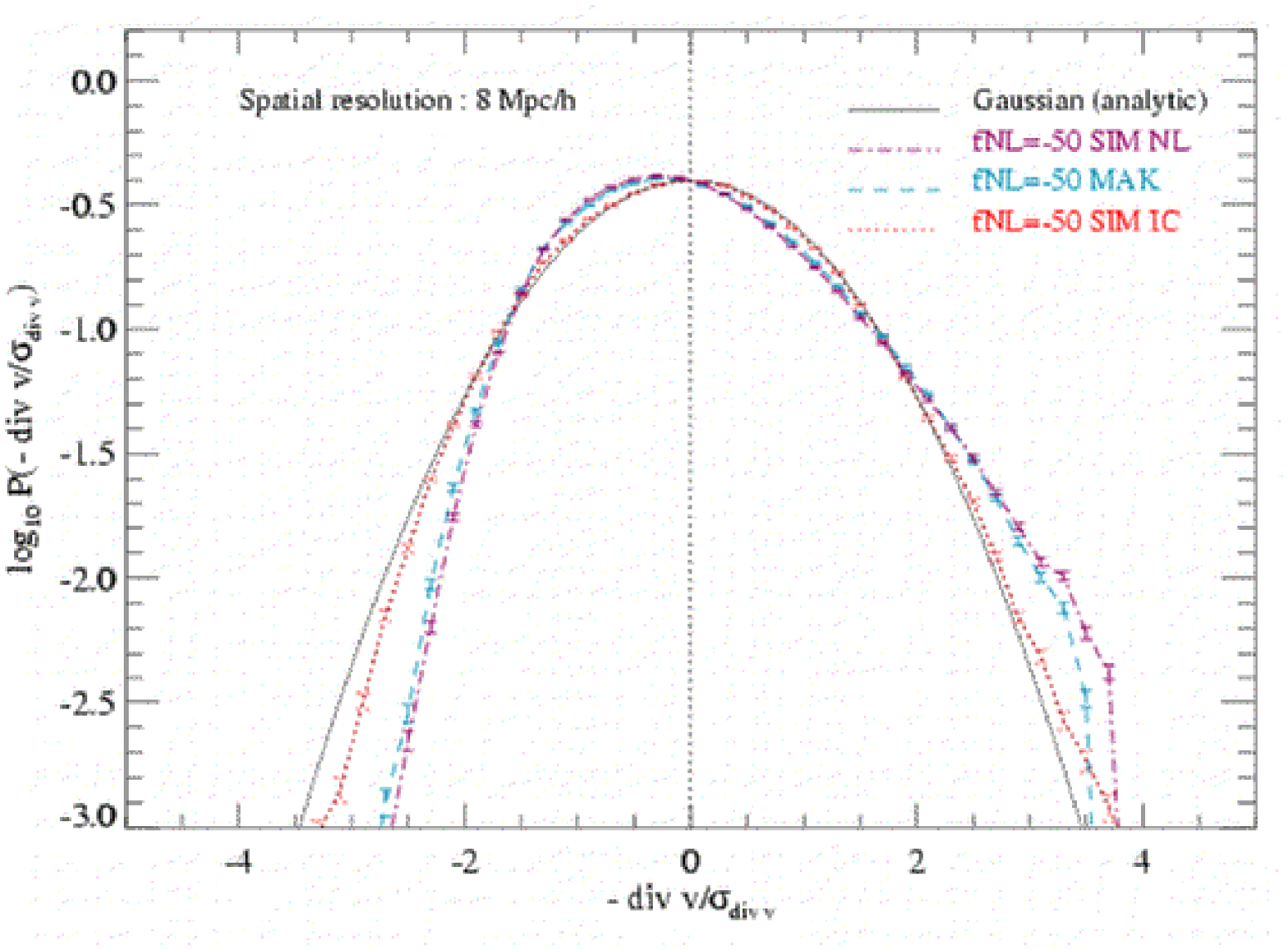}
& 
\includegraphics[width=8cm]{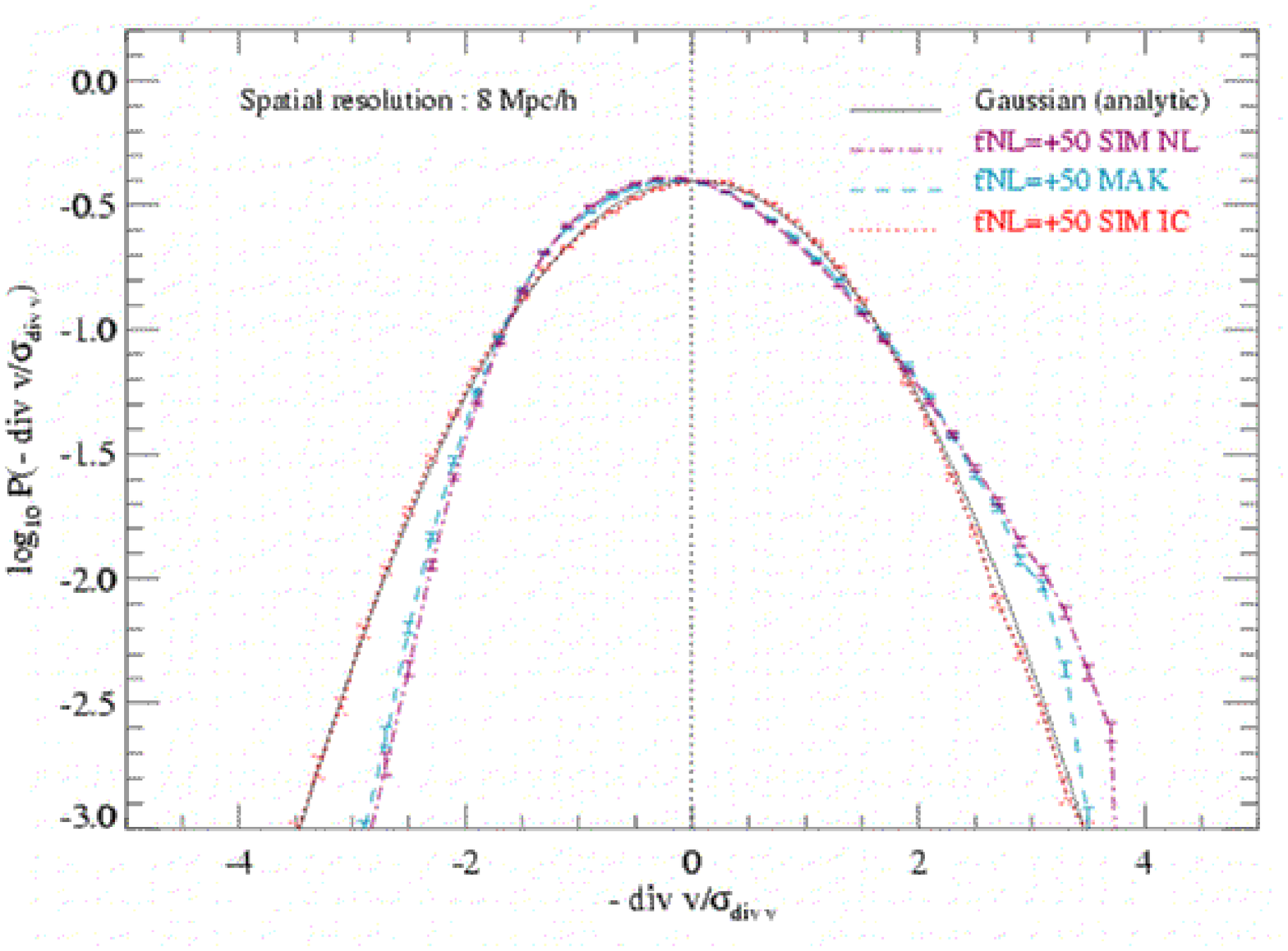} \\
\end{tabular}
\caption{Same as Fig.~\ref{fig:dens_PDF_Gauss}, but for the series of
  quadratic models. $Q_{-100}$, $Q_{+100}$, $Q_{-50}$, $Q_{+50}$
 respectively correspond to the
upper left, upper right, lower left and lower right panels.}
\label{fig:dens_PDF_Q}
\end{figure*}

Fig.~\ref{fig:dens_PDF_Q} shows the PDF of $-\bf{\nabla\cdot v}$ 
for the SIM IC, MAK and SIM NL, with a top-hat smoothing of 8 \hMpc for the series of 
quadratic models $Q$  (recall that these are models with very small non-Gaussianity).
These models evolve to very similar 
nonlinear stages since they are all weakly non-Gaussian: gravity dominates significantly over
initial non-Gaussianity suggesting that it will be nontrivial to recover the latter.
Once again, 
the reconstructed nonlinear displacement field matches the simulated one for all of these models.
As expected, the fraction of ``ideally'' reconstructed particles is similar to 
that in the Gaussian model ($41-43\%$ versus $37\%$, see last column of Table 1).

\section{Separation of the initial and gravitational-induced 
non-Gaussianities: the spherical collapse model }
\label{sec:Prospects}

We see that MAK reconstructs the {\em nonlinear} displacement field 
with a tremendous accuracy. The problem is that, in order to constrain the
primordial non-Gaussianity, it is essential to recover
the initial density contrast, {\it i.e.} the divergence 
of the {\em linear} displacement field. To achieve this goal it is necessary
to model the effect of gravitational instability on the displacement field.
Several approaches might be chosen: 

\begin{enumerate}
\item[(i)] {\it Semi-analytic field reconstruction method in an iterative way.}
The method would inspire closely from the ideas developped in the
building of ZTRACE reconstruction (Monaco \& Efstathiou 1999), i.e. would 
be comparable to Peak-Patch \citep{bond96} 
or Pinocchio \citep{monaco02}, but in a reverse way. 
More explicitely, the idea is to use a realistic approximation of the dynamics depending
in a unique way on the initial displacement field 
and the goal is to reconstruct the latter in an iterative
manner. For instance, one can 
use Lagrangian perturbation theory up to some order, the
latter being determined by the level of non-Gaussianity 
one aims to probe, e.g. the skewness (second
order needed) or the skewness and the kurtosis 
(third order needed) of the initial distribution
function. The subtlety of this approach is that one needs to {\em truncate} the
sought-after initial fluctuations at some scale 
in order to deal appropriately with shell crossings in over-dense regions.
The corresponding smoothing scale depends on 
location, according to the mass scale of the collapsed
object formed at present time. 

If one uses Lagrangian perturbation theory, the nonlinear displacement field is indeed,
for the growing mode, entirely determined as a function of 
the initial one, even if one has
to take into account the adaptive smoothing of initial conditions 
needed for dealing with collapsed regions,
as has just been explained. In this approach, the 
goal would then be  to solve an implicit equation on the initial displacement
field by equating the subsequent theoretical nonlinear displacement field with the reconstructed one
by MAK. To solve this implicit equation, the most 
obvious approach would rely on an iterative method, which
is unfortunately not guaranteed to converge. 

\item[(ii)] {\it Statistical method relying on loop perturbation theory.} 
A less elegant, but simpler
and still powerful method consists of working directly on the PDF of 
the divergence of the reconstructed displacement field
to infer the PDF of the initial one. Again, some modelling of 
the dynamics is needed, to infer the mapping
between the former and the latter. The best way to achieve that
relies on perturbation theory and its one-loop (or higher order) corrections 
\citep[e.g.,][]{sf96}. The advantage of this approach is that it 
allows one to probe scales which
are close to the nonlinear regime. Since MAK reconstruction 
is very good, even at mildly nonlinear scales,
this method would be optimal, but still quite involved from the
analytic point of view. 

\item[(iii)] {\it Statistical method relying on the top-hat
spherical model.} To simplify furthermore the approach, with
still in mind the goal of  remapping of the probability distribution,
one can use the top-hat spherical model \citep[e.g.,][]{Fos98a,Fos98b}.
This approximation  has been demonstrated to work very well, both from 
the theoretical \citep{Bern92,Bern94a,Bern94b,Gaz98b,Fos98a,Fos98b}
and the numerical points of view \citep{Fos98a,Fos98b}. However, it works 
only in the regime where fluctuations of the considered field (here the
divergence of the displacement) are very small. Indeed, this model presents
wrong loop corrections, but has the considerable advantage of relying 
on a very simple formalism.
\end{enumerate}
It would be beyond the scope of this paper to try to apply approaches (i) and (ii) to our data, 
because they are both rather involved, from the numerical and analytical points of view. 
Instead, we concentrate on the spherical top-hat model, to demonstrate
that it is possible to estimate the level of non-Gaussianity of the initial 
displacement field from the MAK-reconstructed one.
 
Under the top-hat spherical approximation, the divergence of 
the displacement field can be expressed as a function of the density contrast as follows:
\begin{equation}
\psi \equiv \nabla_q {\bf v}=3 \left[ (1+\delta)^{-1/3}-1 \right].
\end{equation}
We now relate $\psi$ to the initial linear density contrast, 
$\psi_L=-\delta_L$. In the spherical top-hat framework,
the approximation
\begin{equation}
1+\delta \simeq \left( 1- \frac{2}{3} \delta_L \right)^{-3/2}.
\end{equation}
relates $\delta$ to $\delta_L$ \citep{Bern94b}.
In the above equation, the time dependence of the growing mode extrapolated
to present time has been included in $\delta_L$. This approximation
turns out to be excellent, independently of the value of the cosmological
parameters \citep{Bern94b}.

We thus obtain
\begin{equation}
\psi=3 \left[ \left(1+\frac{2}{3} \psi_L \right)^{1/2} -1 \right].
\label{eq:simple}
\end{equation}
In the spherical approximation framework, this gives us, in Lagrangian
space, the transformation
from the initial divergence of the displacement field to the final one.  
This formula is unchanged if top-hat smoothing is applied to the 
fields, here $\psi_L$ and $\psi$, under consideration \citep{Bern94a}.

Inverting Eq.~(\ref{eq:simple}), we obtain the simple relation 
\begin{equation}
\psi_L=\frac{3}{2} \left[ \left( 1+ \frac{\psi}{3} \right)^2 -1\right].
\label{eq:notsimple}
\end{equation}

The spherical collapse 
approximation has already been used to explicitly calculate the PDF in a local Lagrangian formalism 
(Protogeros \& Scherrer 1997; Protogeros et al. 1997).  
The above equation gives us the change of variable to obtain the PDF of the 
initial divergence of the displacement field as a function of the 
non-linear, reconstructed one, by simply using
\begin{equation}
P(\psi_L) d\psi_L= P(\psi) d\psi.
\end{equation}
Note that the relation between $\psi$ and $\psi_L$ [Eq.~(\ref{eq:simple})]
imposes an upper bound on the possible values of $\psi_L$, $\psi_L \geq -3/2$, which
in turns implies that $\psi \geq -3$, as already discussed extensively in 
Section~\ref{sec:ResultsGauss:Dens}. This also shows the limits of the spherical collapse
approximation: it is expected to fail for collapsed objects, i.e. 
when $\psi$ approaches the bound $\psi=-3$,  
as illustrated by Fig.~\ref{fig:ScatterSmooth_Gauss}. There is in fact certainly a way to
fix this by using other approximations in the highly nonlinear regime that we do not
explore here (e.g. using the so called ``halo model'' for instance discussed in 
\citealt{Sco01}).

Here we test the spherical collapse model approximation applied to the divergence of the 
MAK-reconstructed displacement field at a scale of $12 \hMpc$, where we know 
that one-loop corrections are expected to be negligible 
and thus where the spherical collapse model
should perform well, even in the non-Gaussian case, as argued by \cite{Fos98b}.

\begin{figure*}
\begin{tabular}{@{}cc@{}}
\includegraphics[width=8cm]{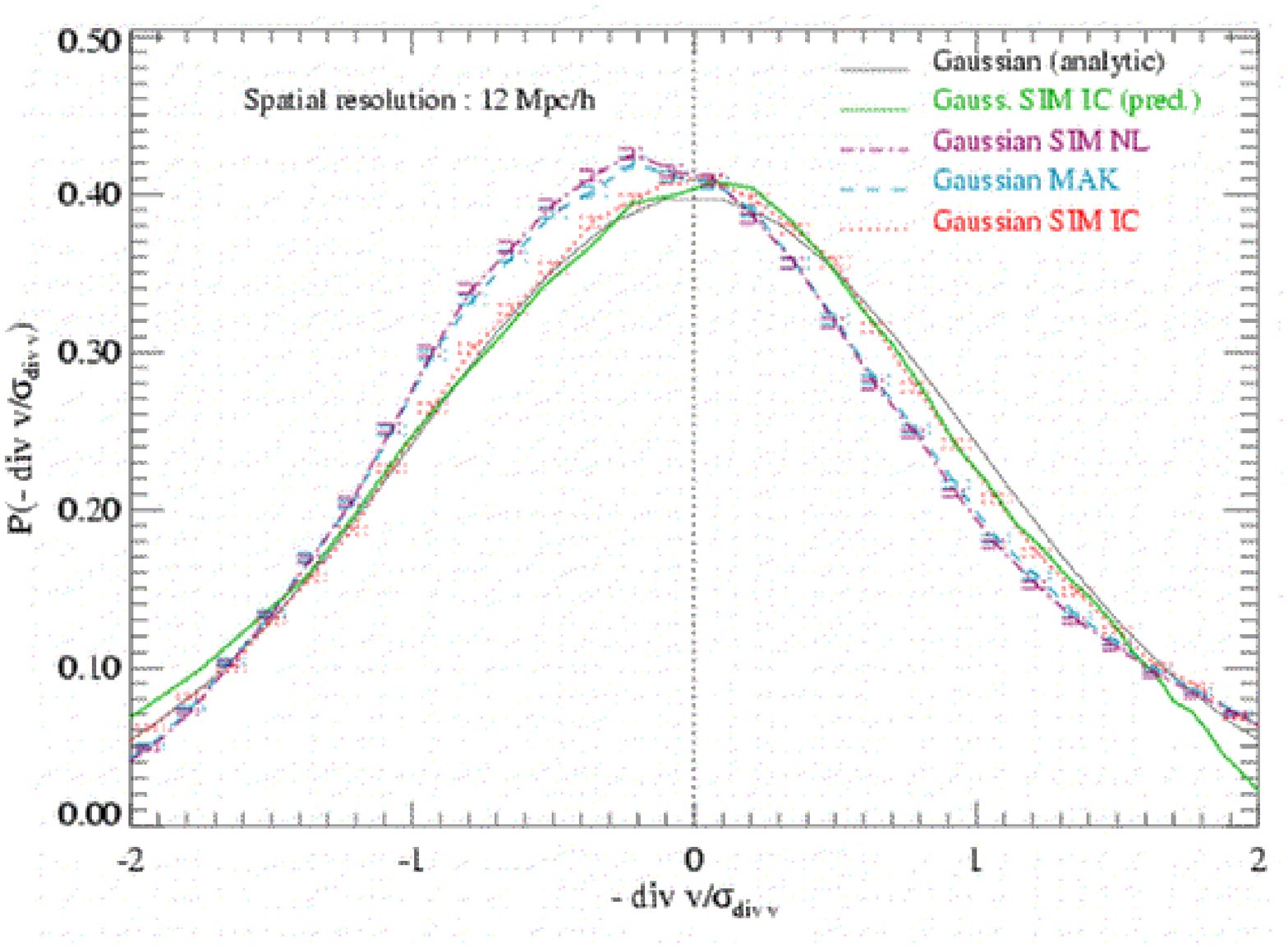}
& 
\includegraphics[width=8cm]{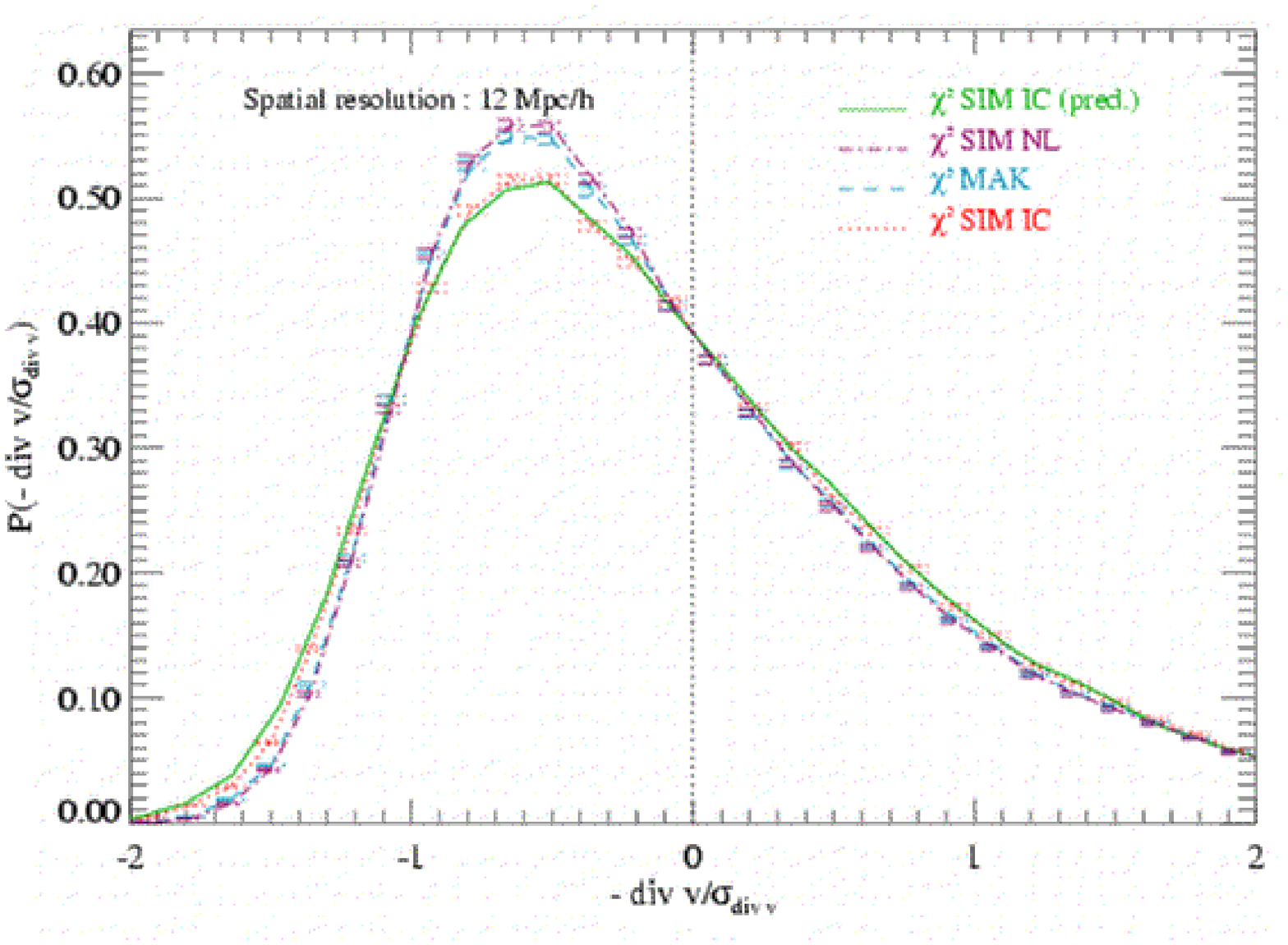} \\
\includegraphics[width=8cm]{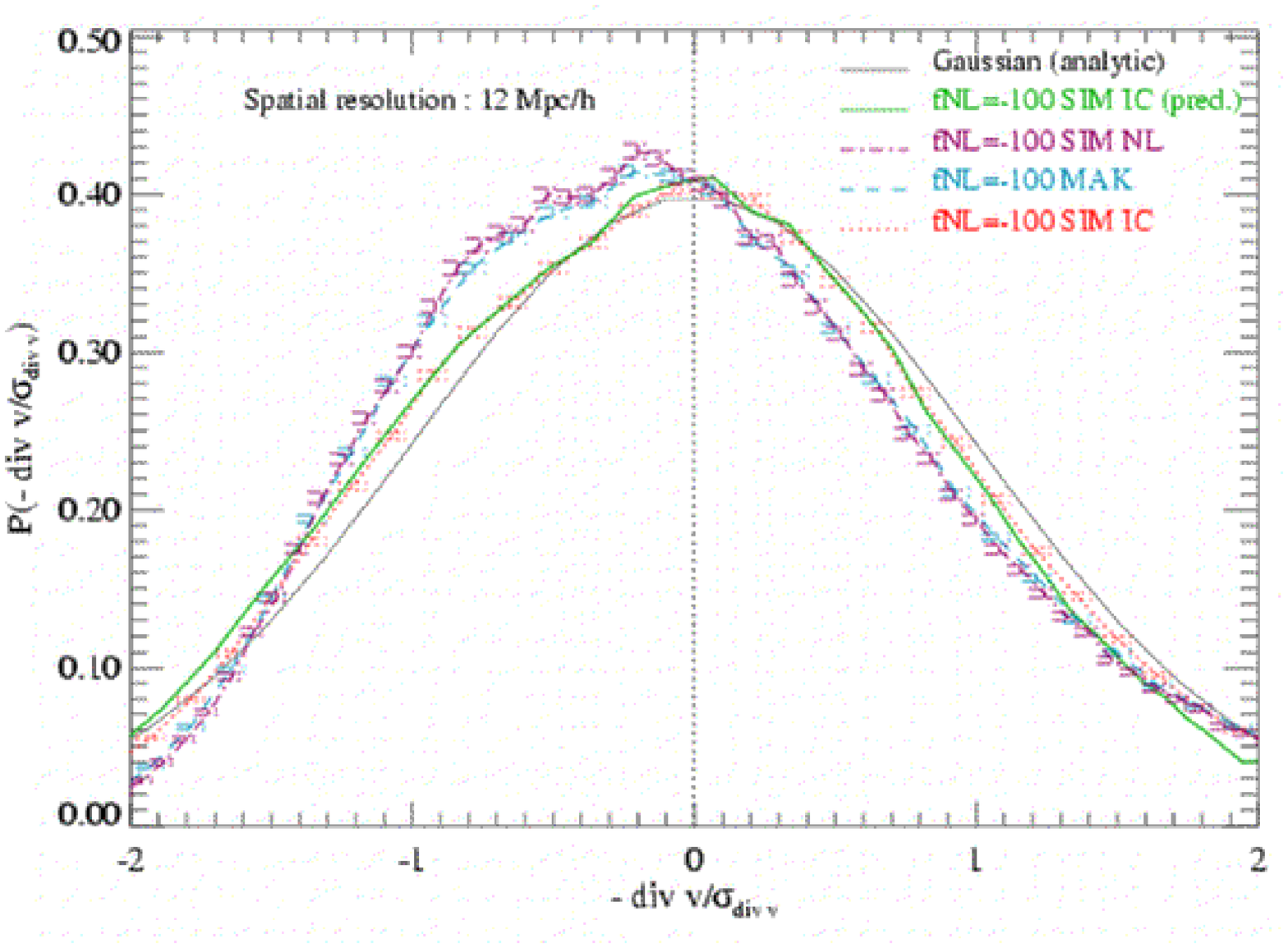}
& 
\includegraphics[width=8cm]{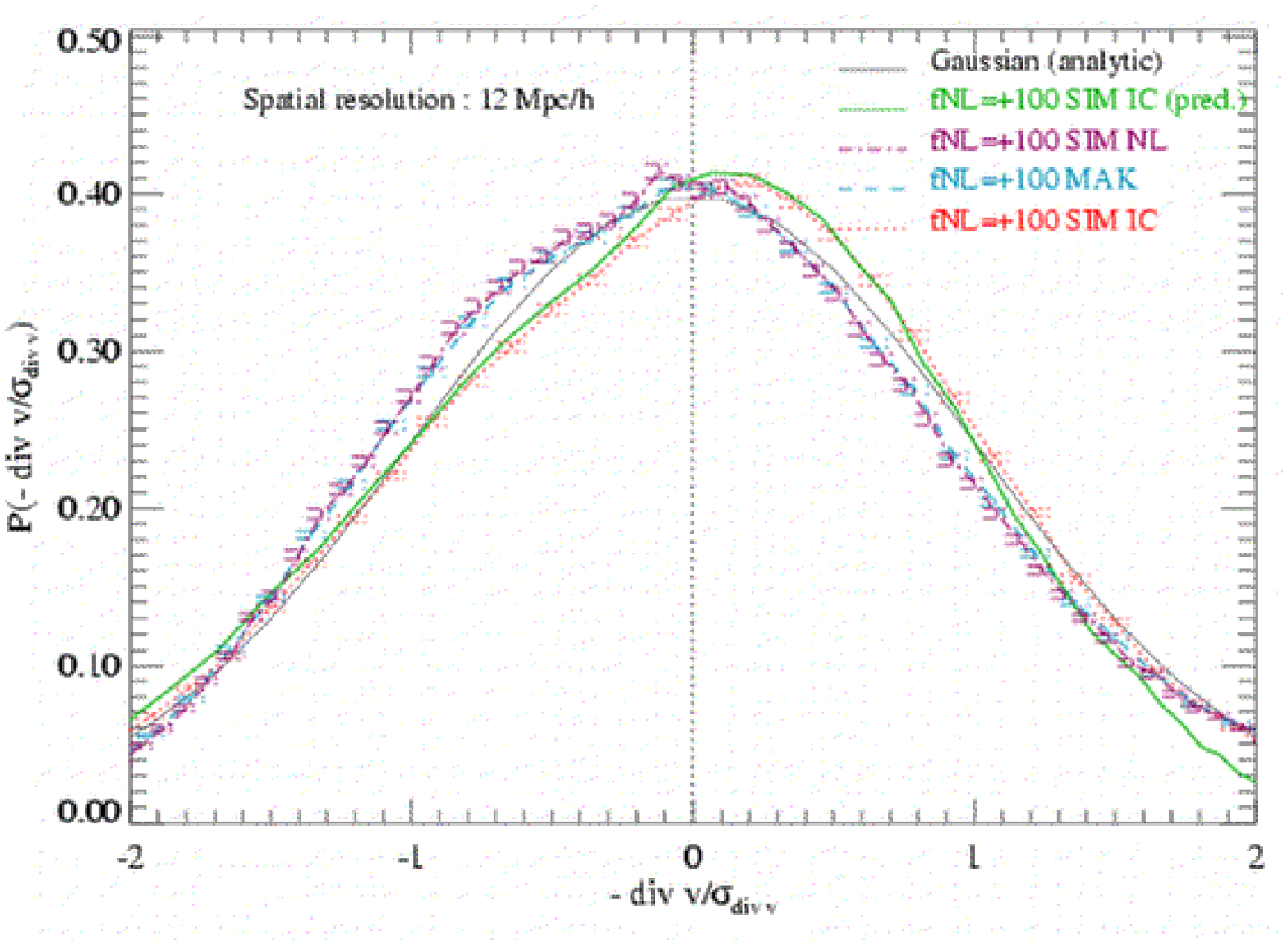} \\
\includegraphics[width=8cm]{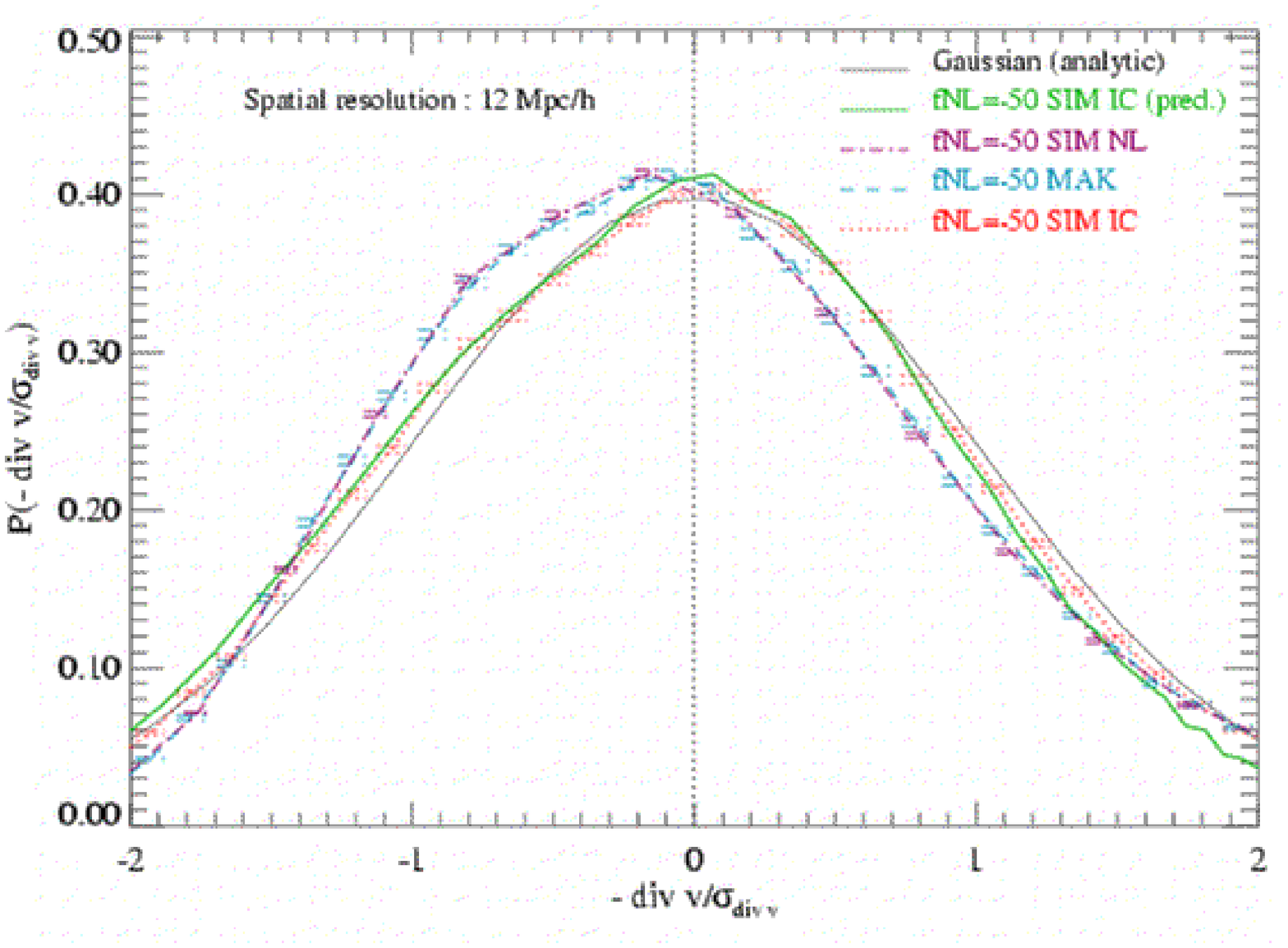}
& 
\includegraphics[width=8cm]{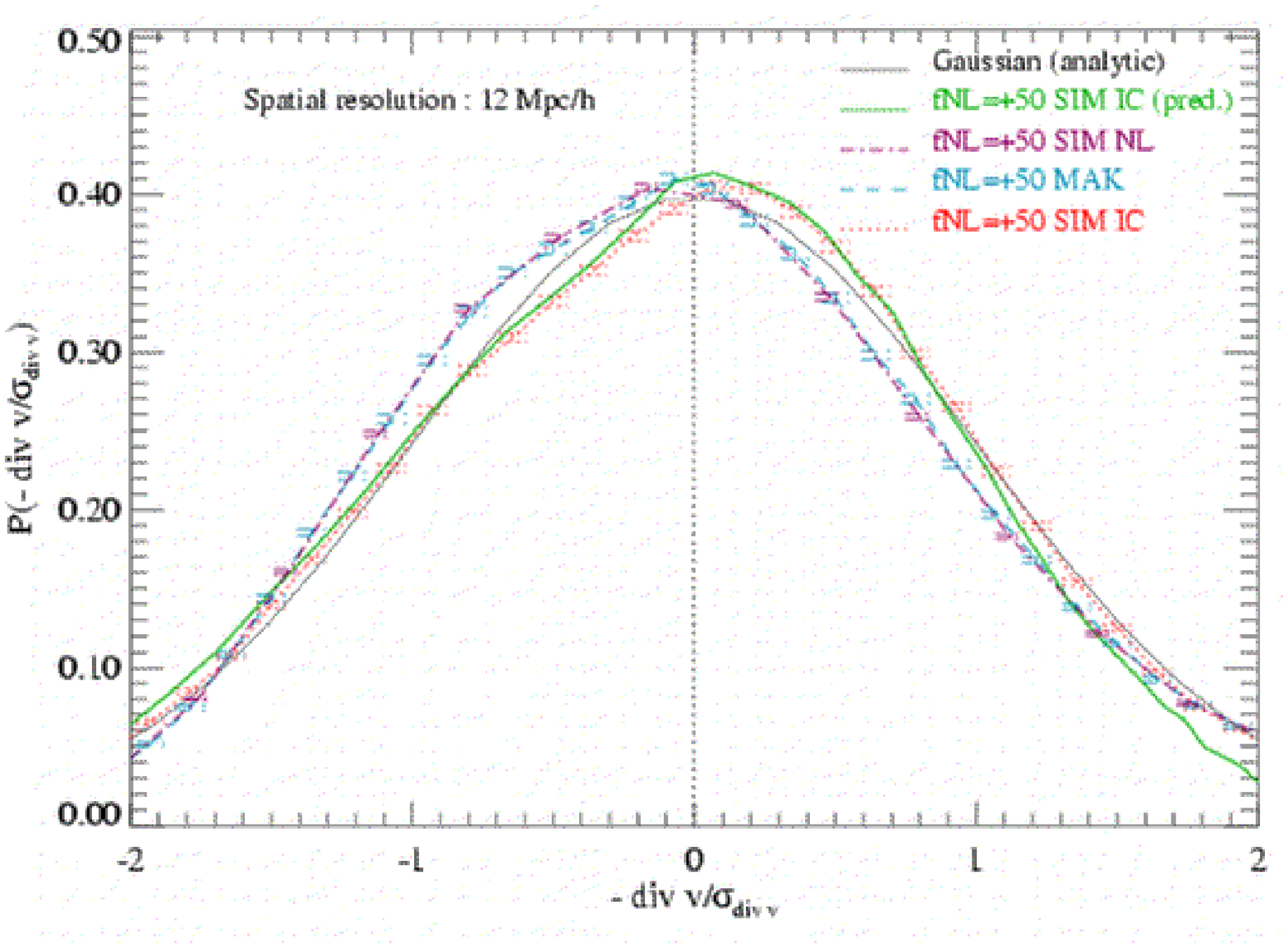} \\
\end{tabular}
\caption[]{Comparison between the PDF of the true initial $-\nabla\cdot{\bf v}$ (SIM IC, dots) to
the one inferred from the reconstructed displacement field using the top-hat spherical
collapse approximation (SIM IC, predicted, thick solid curve), after a smoothing
with a top-hat window of radius $12 \hMpcDot$ The choice of $12 \hMpc$ instead
of $8 \hMpc$ is justified by the plots of Fig. \ref{fig:ScatterSmooth_Gauss}.
There are additional curves as detailed in the panels:
the PDFs of $-\nabla_q\cdot{\bf v}$ for the nonlinear displacement
fields, measured in the simulation (SIM NL, dot dashes) and reconstructed (MAK, dashes), and
finally the pure Gaussian prediction (Gaussian, analytic, thin solid curve, not present on
the upper right panel). Each panel corresponds to a given model as indicated on the figures. 

The important curves to examine on each panel are the dotted ones (initial PDF) and the thick solid ones
(predicted initial PDF from the MAK nonlinear PDF), while keeping the Gaussian limit in
mind: they should in principle superpose if the combination MAK+spherical collapse model works.
This is indeed the case in all panels, except in the tails, especially the right-hand side one, but
only for rather large values of $|-\nabla_q\cdot{\bf v}/\sigma_{\nabla\cdot{\bf v}}| \sim 2$. Indeed,
in this regime, especially for positive values of $-\nabla_q\cdot{\bf v}$, the spherical collapse model
is known to have limitations \citep[e.g.,][]{Bern94b}. Except for 
this, the agreement between the prediction
and the measurement is excellent. It is in fact so good, that 
for the $Q$ models, the weakly non-Gaussian
features are recovered, if one examines the upper part 
of the curves displayed in the four lower panels. The
curves displayed on each are however 
slightly irregular, because the scale considered here is a significant
fraction of the box size and the measured PDFs are 
thus expected to be contaminated by finite volume effects.
This explains for instance the fact that in 
the Gaussian case, the dotted curve does not exactly superposed
to the thin solid curve, i.e. the pure Gaussian 
prediction. Note that even this small deviation is reproduced
by the MAK+top-hat prediction, at least for the right-hand part of the bell-shaped curves.}
\label{fig:SC_pred_linear}
\end{figure*}

We first examine in Fig.~\ref{fig:SC_pred_linear} the Gaussian, $\chi^2$ and $Q$ models. We
shall pay attention to the special case of the PVM model at the end of this section. 
The success of the spherical model is unquestionable. Except in the tails (especially the
right-handed one, as argued above), the shape of the initial PDF is recovered to very
good accuracy. As discussed extensively in the caption of Fig.~\ref{fig:SC_pred_linear},
even subtle features such as those observed in the $Q$ models are reproduced,
lending credence to the fact that it seems possible to detect even small levels
of primordial non-Gaussianity, at least with MAK reconstruction. Of course, this assertion
does not take into account the instrumental and observational effects as shall be discussed later in
the conclusion. Furthermore, our approach here is rather phenomenological, since we do not
present a serious treatment of the statistical significance of the comparison. 

\begin{figure*}
\includegraphics[width=8cm]{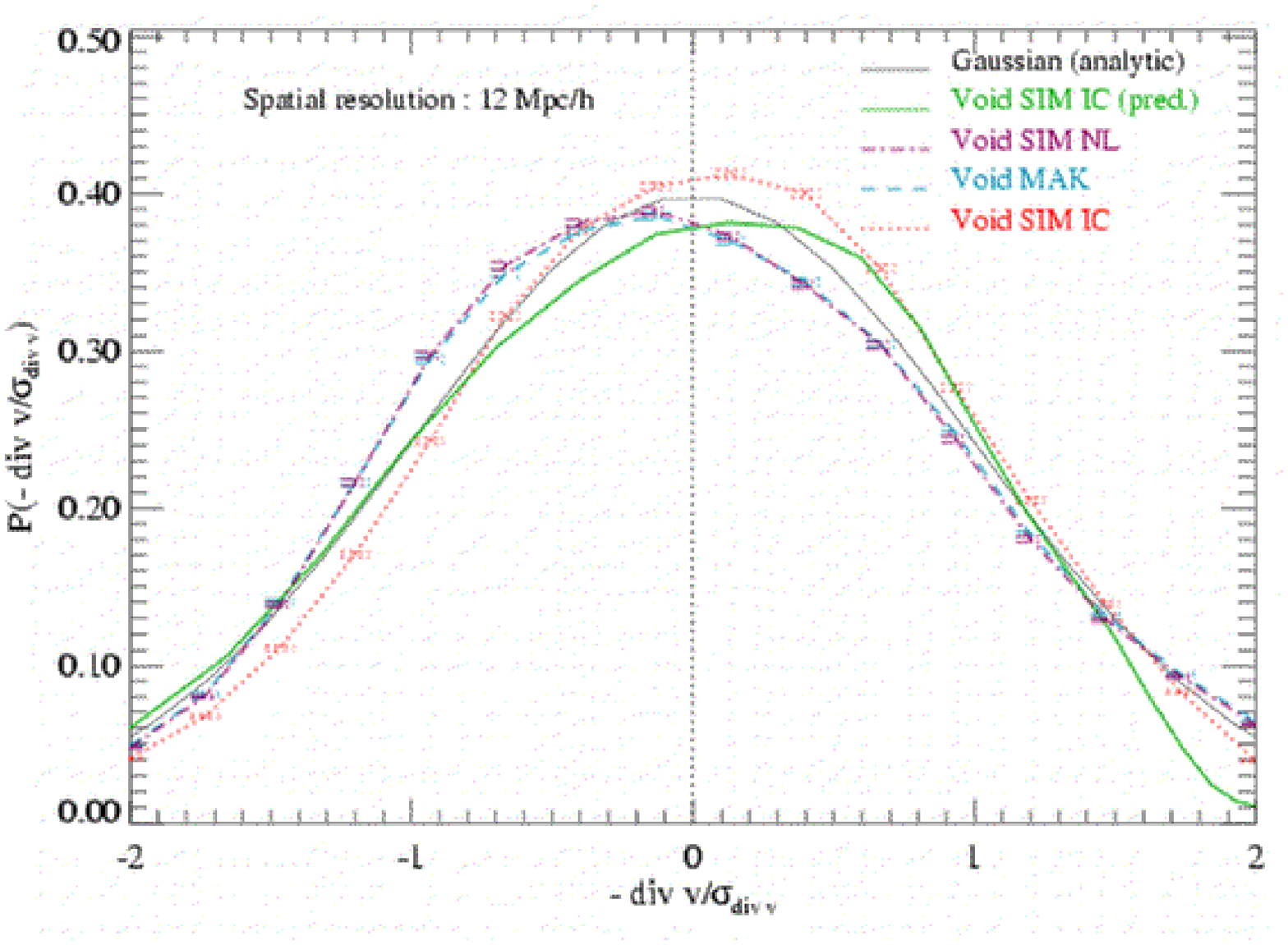}
\caption{Same as in Fig.~\ref{fig:SC_pred_linear} but for the  Void model.}
\label{fig:SC_Void}
\end{figure*}

We now turn to the PVM model. This model deserves a special treatment because the initial
velocity field is not set up according to the Zel'dovich approximation, as explained in
detail in Section~\ref{sec:Models:PVM}. As a consequence, there
are transients even for the variance of the displacement field: in the linear regime, 
the expanding voids are in fact pure transients, but they induce strong density contrasts that
increase the variance of the fluctuations, thus the variance of the divergence of the displacement
field. A consequence is that standard linear theory, with only the growing mode, does
not apply. For instance, the value of $\sigma$ (where $\sigma^2$ is the variance of
the field) at 
$12 \hMpc$ measured in the initial $-\nabla\cdot{\bf v}$ scaled
to present time is $0.94$, larger than the measured one from the nonlinear displacement field, 0.68. 
The spherical collapse model captures only the growing modes: they are determined
by the initial Gaussian field prior to the void seeding. This one had a value
of $\sigma$ of $0.71$, 
hence of the same order as the present one, as expected. In Fig.~\ref{fig:SC_Void}, we thus assumed
for the comparison between true initial $-\nabla\cdot{\bf v}$ PDF and predicted one
a value of $\sigma$ of $0.71$ for the 
latter. In practice, note that the test for non-Gaussianity does not need
the precise knowledge of this initial $\sigma$: only the shape of the PDF matters, and
has to be compared with the best fitting Gaussian in a trustworthy range.

Here the match between the predicted and the 
initial PDF is not so good anymore, although some non-Gaussian features seem to be captured,
namely the right hand excess compared to the Gaussian case, but this might as well be
a mere coincidence. The problem here is that the arguments in favour of the spherical
top-hat model \citep[e.g.,][]{Fos98a,Fos98b} 
might not apply to the PVM model and its very strong transient nature, despite
the fact that the voids are chosen here to be perfectly spherical.\footnote{but they fragment 
in a nontrivial way due to the additional fluctuations induced by the Gaussian field.} The validity of
Eq.~(\ref{eq:simple}) and especially of the assertion that smoothing does not affect this equation
is certainly questionable for the PVM model.

\section{Discussion}
\label{sec:CCL}

In this work, we have tested the Monge-Amp\`ere-Kantorovich (MAK)
reconstruction against $N$-body simulations. We have examined the 
standard $\Lambda$CDM model and 6 additional models 
with non-Gaussian initial conditions: a $\chi^2$ model
where the initial density field is the square of a Gaussian, a model with
primordial voids, and four mildly non-Gaussian models where 
the primordial gravitational potentials includes a small, quadratic correction.

The main results of this paper are as follows:
\begin{itemize}

\item
In its essence, MAK is supposed to reconstruct the nonlinear displacement field between initial
and final positions. The analyses of this paper show that it achieves this with an
unprecedentedly high degree of accuracy, at least at scales larger than the
size of rich clusters, {\it i.e.,} scales above a few Mpc. In particular, it captures in a
nontrivial way the nonlinear contribution from gravitational instability, well
beyond the Zel'dovich approximation.

\item
From the MAK-reconstructed nonlinear displacement field, one can infer the
present velocity field, using the Zel'dovich approximation. Again, our $N$-body
simulation analyses demonstrate the success of MAK in fulfilling this goal.

\item
The displacement field provided by MAK can be used as well to constrain the
statistical nature of the primordial density field. But, since MAK encodes
nonlinearities due to gravitational clustering, it is difficult to disentangle
these dynamical contributions from possible initial non-Gaussianities.
However, as illustrated here in a simple way with the spherical collapse model, 
we show that it is possible to truly recover the statistical properties of the 
primordial, linear density field using additional modeling of the dynamics.
Furthermore, we envisage possible improvements to be made by 
elaboration of MAK to a more sophisticated modeling of the dynamics
combined with adaptive smoothing
of the initial fluctuations [similarly as in ZTRACE (Monaco \& Efstathiou
1999)].

\end{itemize}

The results presented in this paper are obtained in 
the ``idealized" framework of $N$-body simulations.
Extra complications arise in reconstruction from real galaxy 
catalogues (for application of MAK to real galaxy 
catalogues see \citealt{Moh04,Moh05}). Here, we discuss  
redshift space distortion, edge-effects, biasing and catalog incompleteness,
which are the most relevant issues for MAK reconstruction: 

\begin{enumerate}

\item{\it Redshift space distortion:}
MAK has already been directly applied to redshift catalogues by modifying the cost 
function~(\ref{eq:assign}) using once again the Zel'dovich 
approximation (for detail, see~\citealt{Moh04}).
An alternative method 
would be to deduce real space positions from redshift space data, using MAK 
in an iterative way. 
Indeed, we have shown in
paragraph~\ref{sec:ResultsGauss:PecVels} very 
good agreements between the present-day simulated peculiar velocities and 
the MAK-reconstructed ones already on $\sim$ 8 \hMpc scales. 

\item {\it Boundary effects:}
in this paper, we have partly addressed edge and tidal effects using 
the dense samples. We have noticed that these effects
are small but the volume considered was large, a cube 100 \hMpc
aside. Clearly, large scale tidal effects and wrong assignments near the
edges of a catalog would be important if it had a small volume coverage
or/and intricate boundaries. 

\item{\it Biasing:}
in a real galaxy catalog, there is the problem of biasing that is the
relationship between the present total matter distribution and the present
observed light distribution. For each observed galaxy of a given luminosity, $L$,
one has to infer a given mass which for instance can be a simple function of
$L$. In this way, one can transform a distribution of points with given
luminosities to a distribution of points with given masses. A supplementary
complication arises since MAK requires in essence 
all the points to have the same mass. To achieve this, one can break the
galaxies into equal mass particles,
place them at the position of the galaxy and perform MAK
reconstruction.\footnote{In practice, one distributes the particles
into Gaussian clouds with negligible dispersion around the original galaxy. One
reason for this is to speed up the assignment algorithm.}
In the hierarchical clustering framework, galaxies
are located in dark matter halos, that is in condensed objects.
We know from the results of this paper that MAK reconstructs very well
the nonlinear displacement field even for regions that have experienced
shell-crossings. In the latter case, the only problem is the wrong assignment of the
particles inside the collapsed objects, but this does not affect the reconstruction of
the displacement field significantly, and hence the velocity field inferred from it using
the Zel'dovich approximation. Therefore, provided that the right mass has been given to each
galaxy and that the catalog is complete enough so that the underlying total 
mass distribution is well-traced, the reconstruction 
is expected to perform well~(\citealt{Moh04,Moh05}).  There is however a subtlety which has
to be taken into account, during mass assignment to galaxies: a different
treatment has  to be performed for field galaxies which have their own
halo and galaxies belonging to rich clusters. In the latter case, what matters 
is the mass of the halo of a cluster and not the mass of the ``sub-halos" of
each of its galaxies~(\citealt{Moh05}).

\item{\it Luminosity segregation:}
the last issue to consider is catalog incompleteness. For instance, in a
standard magnitude-limited catalog, there is a luminosity segregation effect
which arises due to fewer and brighter objects with increasing distance from
the observer. It is possible to compensate for it by
assigning a larger mass to more distant 
galaxies~(\citealt{Moh05}): this means that two galaxies of
the same luminosity but at different distances from the observer will be given
different masses. With this procedure, the reconstruction is in principle expected to be 
as least biased as possible. However, clearly the signal-to-noise ratio
would decrease with distance from the observer, and more subtlely, the expected strong dependence
of galaxy clustering with luminosity might complicate the interpretation of the results:
obviously, it is needed to have catalogs complete enough in the faint end of the luminosity
function in order to treat appropriately the biasing issues discussed in previous point.
\end{enumerate}


\section*{Acknowledgements}

We thank Michel H\'enon for providing us with the fast
cosmologically-adapted dense and sparse versions of auction
code for solving the assignment problem. We also thank Francis Bernardeau for important 
insights into spherical collapse model. RM was supported by a Marie
Curie HPMF-CT 2002-01532 and a European Gravitational Observatory
(EGO) fellowship at 
the school of astronomy of the university 
of Cardiff, UK. HM acknowledges financial support from
a UK PPARC fellowship. 
Special thanks go to Jacques Colin and Uriel Frisch for invaluable support at 
the Observatoire de la C\^ote d'Azur
where the major part of this work was carried out.



\bsp
\label{lastpage}

\bibliographystyle{mn2e}
\bibliography{Citations}

\end{document}